\documentclass[fleqn,usenatbib]{mnras}

\usepackage[T1]{fontenc}

\DeclareRobustCommand{\VAN}[3]{#2}
\let\VANthebibliography\thebibliography
\def\thebibliography{\DeclareRobustCommand{\VAN}[3]{##3}\VANthebibliography}

\usepackage{graphicx}	
\usepackage{amsmath}	
\usepackage{amssymb}	
\usepackage{upgreek}

\title[Broadband radio study of PSR~J0250+5854]{A broadband radio study of PSR~J0250+5854: the slowest-spinning radio pulsar known}

\author[C. H. Agar et al.]{
C.~H. Agar,$^{1}$\thanks{E-mail: crispinagar.work@gmail.com}
P. Weltevrede,$^{1}$ 
L. Bondonneau,$^{2,3}$ 
J.-M. Grie{\ss}meier,$^{2,4}$ 
J.~W.~T. Hessels,$^{5,6}$ 
\newauthor
W.~J. Huang,$^{7,8,9}$ 
A. Karastergiou,$^{10}$ 
M.~J. Keith,$^{1}$ 
V.~I. Kondratiev,$^{5,11}$ 
J. K\"{u}nsem\"{o}ller,$^{12}$
\newauthor
D. Li, $^{13,14,15}$ 
B. Peng,$^{13}$ 
C. Sobey,$^{16}$ 
B.~W. Stappers,$^{1}$ 
C.~M. Tan,$^{17,18}$ 
G. Theureau,$^{2,4,19}$ 
\newauthor
H.~G. Wang,$^{7,8,9}$ 
C.~M. Zhang,$^{13,14}$ 
B. Cecconi,$^{3,4}$ 
J.~N. Girard,$^{20}$ 
A. Loh,$^{3}$ 
and P. Zarka$^{4,21}$ 
\\
$^{1}$Jodrell Bank Centre for Astrophysics, School of Physics and Astronomy, University of Manchester, Manchester M13 9PL, UK\\
$^{2}$Laboratoire de Physique et Chimie de l'Environnement et de l’Espace (LPC2E) Universit\'{e} d’Orl\'{e}ans/CNRS, Orl\'{e}ans, France\\
$^{3}$LESIA, Observatoire de Paris, CNRS, PSL, SU/UP/UO, 92195 Meudon, France\\
$^{4}$Station de Radioastronomie de Nan\c{c}ay, Observatoire de Paris, PSL Research University, CNRS, Univ. Orl\'{e}ans, OSUC, 18330 Nan\c{c}ay, France\\
$^{5}$ASTRON, Netherlands Institute for Radio Astronomy, Oude Hoogeveensedijk 4, 7991 PD Dwingeloo, The Netherlands\\
$^{6}$Anton Pannekoek Institute for Astronomy, University of Amsterdam, Science Park 904, 1098 XH, Amsterdam, The Netherlands\\
$^{7}$Department of Astronomy, School of Physics and Materials Science, Guangzhou University, Guangzhou 510006, China\\
$^{8}$Astronomical Science and Technology Research Laboratory of Department of Education of Guangdong  Province, Guangzhou 510006, China\\
$^{9}$Xinjiang Astronomical Observatory, Chinese Academy of Sciences, 150 Science 1-Street, Urumqi, Xinjiang 830011, China\\
$^{10}$Oxford Astrophysics, Denys Wilkinson Building, Keble Road, Oxford, OX1 3RH, UK.\\
$^{11}$Astro Space Centre, Lebedev Physical Institute, Russian Academy of Sciences, Profsoyuznaya Str 84/32, Moscow 117997, Russia\\
$^{12}$Fakult\"{a}t f\"{u}r Physik, Universit\"{a}t Bielefeld, Postfach 100131, 33501 Bielefeld, Germany\\
$^{13}$CAS Key Laboratory of FAST, National Astronomical Observatories, Chinese Academy of  Sciences, Beijing 100101, China \\
$^{14}$University of Chinese Academy of Sciences, Beijing 100049, China\\
$^{15}$NAOC-UKZN Computational Astrophysics Centre, University of KwaZulu-Natal, Durban 4000, South Africa\\
$^{16}$CSIRO Astronomy and Space Science, PO Box 1130, Bentley, WA 6102, Australia\\
$^{17}$Department of Physics, McGill University, 3600 rue University, Montr\'eal, QC H3A 2T8, Canada\\
$^{18}$McGill Space Institute, McGill University, 3550 rue University, Montr\'eal, QC H3A 2A7, Canada\\
$^{19}$Laboratoire Univers et Th\'{e}ories, Observatoire de Paris, Universit\'{e} PSL, CNRS, Universit\'{e} de Paris, 92190 Meudon,France\\
$^{20}$AIM, CEA, CNRS, Universit\'e Paris-Saclay, Universit\'e de Paris, F-91191 Gif-sur-Yvette, France\\
$^{21}$Sorbonne Universit\'e, Universit\'e de Paris, 5 place Jules Janssen, 92195 Meudon, France
}

\date{Accepted XXX. Received YYY; in original form ZZZ}

\pubyear{2021}

\begin{document}
\label{firstpage}
\pagerange{\pageref{firstpage}--\pageref{lastpage}}
\maketitle

\begin{abstract}
We present radio observations of the most slowly rotating known radio pulsar PSR~J0250+5854. With a 23.5~s period, it is close, or even beyond, the $P$-$\dot{P}$ diagram region thought to be occupied by active pulsars. The simultaneous observations with FAST, the Chilbolton and Effelsberg LOFAR international stations, and NenuFAR represent a five-fold increase in the spectral coverage of this object, with the detections at 1250~MHz (FAST) and 57~MHz (NenuFAR) being the highest- and lowest-frequency published respectively to date. We measure a flux density of $4\pm2$~$\upmu$Jy at 1250~MHz and an exceptionally steep spectral index of $-3.5^{+0.2}_{-1.5}$, with a turnover below $\sim$95~MHz. In conjunction with observations of this pulsar with the GBT and the LOFAR Core, we show that the intrinsic profile width increases drastically towards higher frequencies, contrary to the predictions of conventional radius-to-frequency mapping. We examine polarimetric data from FAST and the LOFAR Core and conclude that its polar cap radio emission is produced at an absolute height of several hundreds of kilometres around 1.5~GHz, similar to other rotation-powered pulsars across the population. Its beam is significantly underfilled at lower frequencies, or it narrows because of the disappearance of conal outriders. Finally, the results for PSR~J0250+5854 and other slowly spinning rotation-powered pulsars are contrasted with the radio-detected magnetars. We conclude that magnetars have intrinsically wider radio beams than the slow rotation-powered pulsars, and that consequently the latter's lower beaming fraction is what makes objects such as PSR~J0250+5854 so scarce.

\end{abstract}



\section{Introduction}
\label{sec: introduction}
Pulsars are rapidly rotating, highly magnetised neutron stars; however, some pulsars rotate significantly less rapidly than others. In this paper we discuss observations of PSR~J0250+5854, a radio pulsar with a period $P=23.5$~s discovered by \citet{TB+2018} in the Low Frequency Array (LOFAR) Tied-Array All-Sky Survey \citep[LOTAAS;][]{SC+2019}. It was detected with the LOFAR High Band Antenna array between 110--190~MHz, and with the Green Bank Telescope (GBT) between 300--400~MHz. It is the longest-period radio pulsar discovered to date, more than twice the period of the second slowest-spinning \citep[PSR~J2251$-$3711 at $P = 12.1$~s;][]{MK+2020} and almost three times slower than the well-studied 8.5~s pulsar PSR~J2144$-$3933 \citep{YMJ1999}.
Finding such a slow pulsar is rare, which may be explained by the fact that slower pulsars have lower beaming fractions which make it less likely that they are visible to an observer on Earth, and practical limitations due to selection effects, making them significantly harder to identify in pulsar survey data if only a small number of pulses are present. The presence of red noise in periodicity searches further hinders their identification \citep[e.g.][]{LB+2015,HKR2017}. In addition, slower pulsars lose their ability to create electron-positron pairs and accelerate them sufficiently to produce the detectable coherent radio emission \citep{Sxx1971}. Therefore, models for radio emission can be constrained by finding the slowest pulsars that still remain active. As discussed in \citet{TB+2018}, this means it lies in a relatively empty part of the $P$-$\dot{P}$ diagram, beyond the `death valley' as defined by \citet{CRx1993} and the vacuum-gap curvature radiation death line proposed by \citet{ZHM2000}. However, its continued activity is consistent with the partially screened gap model \citep[e.g.][]{Sxxx2013}. Although PSR~J2144$-$3933 rotates almost three times faster, its much smaller spin-down rate $\dot{P} = 4.96\times10^{-16}$~s~s$^{-1}$ makes it more constraining for death line models \citep{MB+2020}.

In this work we present the first detection of PSR~J0250+5854 at frequencies between 1 and 1.5~GHz using the Five-hundred-metre Aperture Spherical Radio Telescope (FAST), along with simultaneous observations with the UK608 and DE601 LOFAR international stations, and NenuFAR (New Extension in Nan\c{c}ay Upgrading loFAR). The NenuFAR detection at 57~MHz is the lowest frequency detection of PSR~J0250+5854 published to date, and the FAST detection the highest, resulting in an extension by a factor of $\sim$5 in spectral coverage of this unique source in the radio domain. The radio frequency evolution is a key part of understanding the pulsar emission mechanism. This is particularly relevant for pulsars such as PSR~J0250+5854 which are on the cusp of the death line. Multi-wavelength observations can provide information on features including the spectral index (how the flux of the pulsar changes with frequency), and changes in the shape and polarisation properties of the radio beam.

Measurements of pulsar radio spectra from a large population began with \citet{Sxx1973}, \citet{MMx1980} and \citet{IK+1981} using frequencies around and below 100~MHz. Most pulsars were found to have steep spectra which could be modelled with a simple power law $S_\nu \propto \nu^k$, where $S_\nu$ is the mean flux density at some frequency $\nu$ and $k$ is the spectral index. For some pulsars deviations from this relation were identified in the form of a turn-over at low frequencies which can be attributed to absorption mechanisms, whilst others show a cut-off at high frequencies due to a steepening or break in the spectrum \citep{Sxx1973}. 
Recent studies of radio pulsar spectral indices show a reasonably broad distribution centred at approximately $-1.5$ \citep{BK+2016, JS+2018}.

With its period of 23.5~s, PSR~J0250+5854 has an extremely large light-cylinder (with a radius $R_\mathrm{LC} = cP/2\pi =  1.123\times10^{6}$~km, where $c$ is the speed of light), and hence a tiny polar cap which connects to the open field line region. The diameter of the polar cap is $D_\mathrm{PC} \approx 2R\sqrt{R/R_\mathrm{LC}} \approx~60$~m, where $R = 10$~km is the canonical neutron star radius \citep[e.g.][]{Sxx1971}. By comparison, a pulsar with a period $P = 0.5$~s would have $D_\mathrm{PC} \approx 410$~m. This implies that for typical emission heights of hundreds of kilometres, the radio beam of PSR~J0250+5854, and hence the duty cycle of its radio pulse, can be expected to be very narrow. Indeed \citet{TB+2018} reported a pulse width of only $\sim$1$\degr$ at 129, 168, and 350~MHz. The shapes of pulse profiles, even after correcting for propagation effects, are in general observed to be frequency-dependent. Often, the profile width decreases with increasing frequency which suggests that higher-frequency emission is produced lower in the magnetosphere. This correlation is known as radius-to-frequency mapping (RFM hereafter). RFM was first theorised by \citet{RSx1975} -- the magnetospheric electron density, hence plasma frequency, is expected to decrease with increasing altitude, thereby predicting that the radio beam expands with decreasing frequency. A number of pulsars have been found to deviate from this relation \citep[e.g.][]{Txx1991, CWx2014, PH+2016} -- this suggests that not necessarily the same magnetic field lines are active at all frequencies (or emission heights), resulting in the appearance and disappearance of profile components with observing frequency \citep[e.g.][]{Cxx1978, MRx2002}. This can obfuscate the geometrical interpretation of measured profile widths, and is discussed further in Sec.~\ref{sec: discuss width}. Radio polarisation data can help in disentangling these effects, which we further explore for PSR~J0250+5854 in Sec.~\ref{sec: polarisation and geometry}, although degeneracies often remain \citep[e.g.][]{KJ+2010}.

The structure of this paper is as follows. In Section \ref{sec: observations} the new observations are described, followed by a brief explanation of the radio-frequency interference (RFI) excision techniques used. The analysis of the data described in Section \ref{sec: analysis} is divided into three parts: pulse profile evolution with frequency, polarisation properties, and the spectral shape of the pulsar flux density. These results are then discussed in a broader context in Section \ref{sec: discussion}, and our conclusions are summarised in Section~\ref{sec: conclusions}.

\section{Observations}
\label{sec: observations}

As part of a shared-risk proposal, PSR~J0250+5854 was observed on the 22nd May 2019 with FAST, a facility built and operated by the National Astronomical Observatories, Chinese Academy of Sciences \citep{FASTpaper, LW+2018}. With an effective aperture of 300~m in diameter, it is the world's largest single-dish radio telescope, and is located in a natural depression in Guizhou Province. The central beam of the high-performance 19-beam receiver operating between 1 and 1.5~GHz \citep{19beam} was used. Two LOFAR international stations -- UK608 (United Kingdom) and DE601 (Germany) --- and NenuFAR (France) provided overlapping observations. Chilbolton is home to the UK LOFAR station UK608, formally known as the Rawlings Array and the DE601 station is located at Effelsberg\footnote{DE601 was operated as part of the German Long Wavelength (GLOW) consortium at the time of our observations, in a coherently dedispersed folding mode.}. They each consist of two sub-arrays: the High Band Antenna (HBA; 110--240~MHz) and Low Band Antenna (LBA; 10--90~MHz) \citep{lofarPaper, SH+2011}, although only the HBA were used in this project and the bandwidth was limited to 110--190~MHz. NenuFAR at the time of our observations consisted of 52 groups of 19 dual-polarised antennas, operating between 10 and 85~MHz \citep{ZD+2020}. It is located alongside and extends the capabilities of the Nan\c{c}ay LOFAR station (FR606). Table \ref{tab: observations} gives a summary of the overlapping observations conducted. The LOFAR international stations were observing over the full duration of the FAST observations, and in the case of NenuFAR significantly longer.

\begin{table*}
	\centering
	\caption{Observation properties of the analysed observations. There is a full overlap of the data for the period during which FAST was recording data for PSR~J0250+5854 on the 22nd May 2019 (these observations are denoted with an asterisk (*)). `Resolution' is the number of samples (pulse longitude `bins') per pulse period. The GBT and LOFAR Core observations wer performed on 2017-10-25 and 2017-10-28 respectively.}
	\label{tab: observations}
	\begin{tabular}{crrrrrrrc} 
		\hline
	    Observation & Centre freq. & Bandwidth & No. freq. & Start time & No. pulses & Length & Resolution & Full Stokes\\
	    & (MHz) & (MHz) & channels & (UTC) & & (hh:mm:ss) & (No. Bins)&\\
		\hline
		FAST *        & $1250.00$ & $500.0$   & 4096 & 02:34:07  & 100   & 00:39:13  & 8192 & Y\\
		FAST *	    & $1250.00$ & $500.0$   & 4096 & 03:31:23 & 153   & 01:00:01  & 8192 & Y\\
		DE601 *        & $158.55$  & $71.4$    & 488  & 01:56:19  & 417   & 02:43:34  & 1024  & N \\
		UK608 *  & $149.71$  & $95.2$    & 1952 & 02:02:50  & 382   & 02:29:50  & 8192  & N\\
		NenuFAR *     & $56.54$   & $75.0$    & 384  & 02:03:18  & 1528  & 09:59:21  & 2048  & N \\
		\hline
		GBT & 350.00 & 100.0 & 4096 & 11:54:28 & 242 & 01:34:55 & 8192 & N\\
		LOFAR Core & 148.93 & 78.1 & 400 & 23:57:16 & 152 & 00:59:49 & 16384 & Y\\
		\hline
	\end{tabular}
\end{table*}

Prior to observing PSR~J0250+5854, FAST performed a $\sim$15-minute observation of a well-known, bright pulsar PSR~J0139+5814 to validate the set-up of the observing system. A noise diode signal was injected into the FAST multibeam receiver to facilitate polarisation calibration. PSR~J0250+5854 was observed for two consecutive hours, interspersed with noise diode observations. Finally, the BL Lacertae object J0303+472 \citep{VVx2006} was observed for purposes of flux calibration -- this source was chosen due to its proximity to PSR~J0250+5854.

All data were folded and de-dispersed with \textsc{dspsr} \citep{SBx2011} using the ephemeris and dispersion measure (DM) reported in \citet{TB+2018} to form a pulse sequence. De-dispersion was done coherently for the DE601, UK608, and NenuFAR data, and incoherently for the FAST data. Flux and polarisation calibration were done using \textsc{psrchive} \citep{psrchive}. Further processing made use of \textsc{psrsalsa} \citep{psrsalsa}\footnote{\url{https://github.com/weltevrede/psrsalsa}} and is described later.

Although not part of the simultaneous observations, this project also made use of an observation of PSR~J0250+5854 using the LOFAR Core stations conducted on 28 October 2017. This was part of a run of observations conducted by \citet{TB+2018}, but is a different data set to the profiles shown in that paper. This LOFAR Core observation was similarly processed with \textsc{psrchive}.

\subsection{Data cleaning techniques}
\label{sec: cleaning}

PSR~J0250+5854 has a low flux-density which, in combination with its extraordinarily long period, makes the analysis susceptible to radio-frequency interference (RFI) that affects the baseline level during a rotation period.

In all datasets the worst-affected frequency channels were identified and excluded from further analysis. The FAST data were affected by stochastic baseline variations that persisted throughout the observation, with a timescale somewhat larger than the pulsar's duty cycle. These baseline variations were removed using \textsc{psrsalsa} by subtracting sinusoids, up to the 23\textsuperscript{rd} harmonic of the pulse period, plus a constant offset fitted to the off-pulse region for each rotation of the star in each frequency and Stokes parameter independently. These sinusoids have periods which significantly exceed the duty cycle of the pulsar, hence the shapes of the pulses were not affected by this process.

The RFI in the DE601 and UK608 observations was very different, appearing as short, bright, impulsive spikes orders of magnitude brighter than the pulsar signal. An effective approach to mitigation was to iteratively clip the brightest samples for each rotation of the star and each frequency channel individually. The clipping is done conservatively to ensure that the pulsar signal is unaffected. With the worst RFI suppressed, the remaining RFI and baseline variations were reduced using the same method described for the FAST data.

The NenuFAR data were recorded during the commissioning phase of the instrument with a coherent de-dispersion pipeline \citep[LUPPI;][]{LUPPI} operating in single-pulse mode. The observations were folded with \textsc{dspsr} and a polynomial of degree two was subtracted from the baseline of each sub-integration to suppress the effect of bandpass variations. The data from two frequency bands were appended after correcting for the appropriate delay\footnote{The observation was recorded in two sub-bands which have a slight offset when de-dispersed, and so had be aligned manually. This is only necessary for older observations conducted before the ``early science'' phase.}. The frequency-resolved observation was cleaned using a modified version of \textsc{coastguard} \citep{coastguard} and final processing was done with \textsc{psrsalsa} in the same way as with the FAST data.

\section{Analysis and results}
\label{sec: analysis}
\subsection{Profile morphology and width evolution}
\label{sec: profile widths}

\begin{figure}
 \includegraphics[width=\columnwidth]{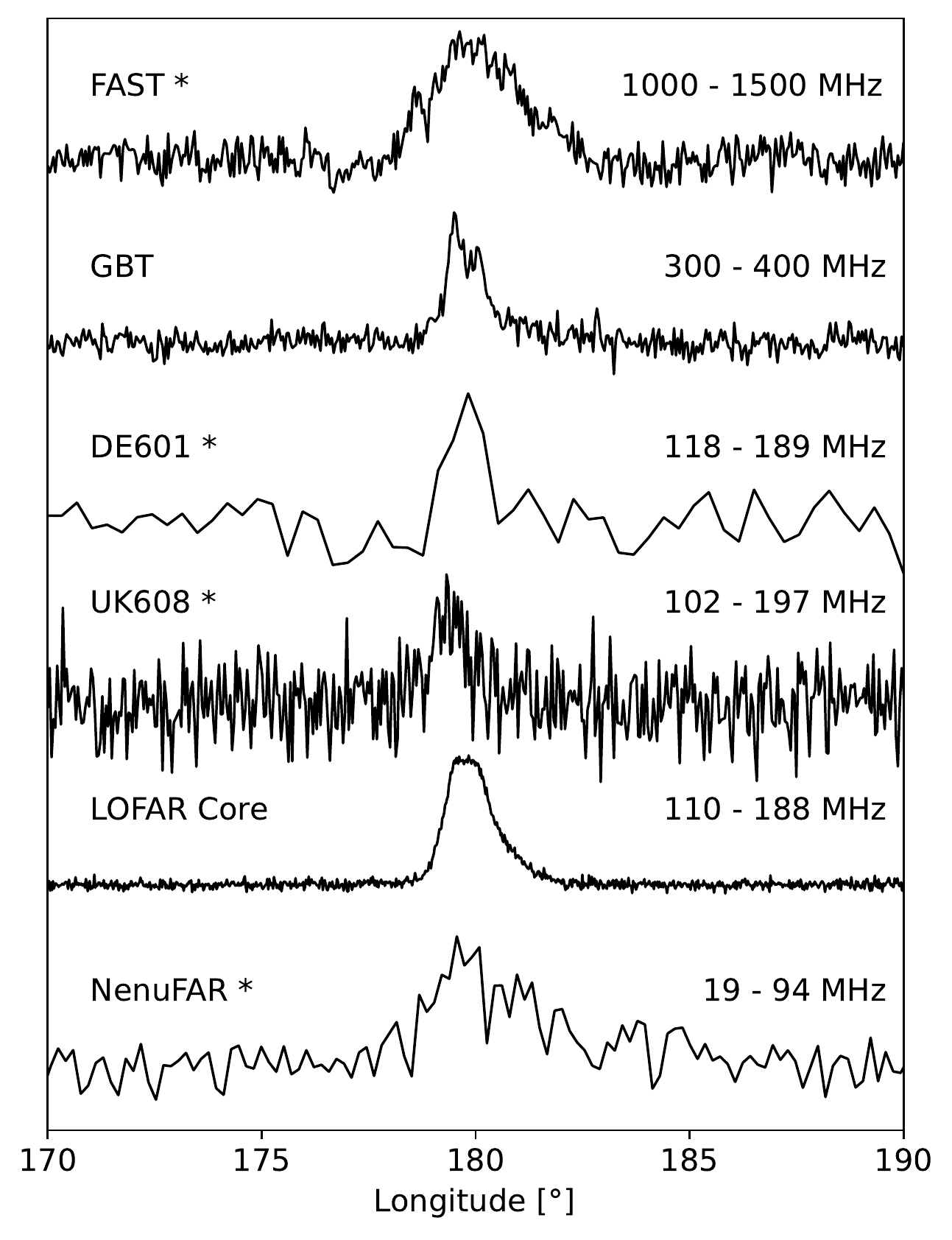}
 \caption{ The pulse profiles of PSR~J0250+5854 at different radio frequencies. The top profile is from FAST (1250~MHz), followed by GBT (350~MHz), then DE601 (154~MHz), UK608 (150~MHz), LOFAR Core (149~MHz), and finally NenuFAR (57~MHz). The FAST, DE601, and UK608 observations are overlapping in time, and are aligned using the known DM and after accounting for geometric delays. The FAST profile is derived from the combination of the two observations. The profiles from the non-simultaneous observations and NenuFAR were visually aligned. The simultaneous observations are denoted with an asterisk (*).}
 \label{fig: profiles}
\end{figure}

Figure \ref{fig: profiles} shows the integrated pulse profile of PSR~J0250+5854 as observed by the four telescopes in order of descending frequency. We also include the profile observed by the GBT at 350~MHz from \citet{TB+2018}, and a LOFAR Core detection at 149~MHz (an observation from 28 October 2017). The profiles for the LOFAR international stations and NenuFAR are obtained from the full-length observation rather than only the overlap period with the FAST observation to increase the signal-to-noise ratio. This is motivated by the fact that there is no evidence that the profile shapes were changing during these observations. Given the large pulse period, a relatively low number of pulses is recorded, making the profiles less stable \citep[e.g.][]{HMTx1975,RRxx1995,LKL+2012}. This is not a significant issue for the following data analysis given the lack of observed variability.

The profiles of the simultaneous observations in Fig.~\ref{fig: profiles} were aligned by correcting for geometric delays associated with the difference in location of the telescopes (taking right ascension to be $02^\mathrm{h}50^\mathrm{m}17\fs78$ and the declination to be $+58\degr54'01\farcs3$ as measured by \citealt{TB+2018}). In addition, the dispersive delay associated with the propagation of the signal through the interstellar medium (ISM) was accounted for by using a DM of $45.281\pm0.003\ \mathrm{cm}^{-3}\mathrm{pc}$ \citep{TB+2018}\footnote{There is no evidence for a change in the DM as the value derived from the NenuFAR data is $45\pm1\ \mathrm{cm}^{-3}\mathrm{pc}$, hence consistent with the DM used.}. The uncertainty on the DM translates to an uncertainty on the dispersion delay between the highest frequency (FAST; 1250 MHz) and lowest frequency (NenuFAR; 57 MHz) observations of around one pulse longitude bin at the highest resolution shown in Fig.~\ref{fig: profiles} (for the FAST and UK608 data). The NenuFAR data was obtained during commissioning phase and so could not be aligned in this way, hence the peak of the profile was aligned visually with the UK608 profile peak.

Only the GBT profile has a clear double-peaked profile morphology. Also the LOFAR Core profile with its flat profile peak, and the asymmetric FAST profile suggest a more complicated profile structure. It is evident that the profile width at frequencies below that of the FAST observation are significantly narrower, opposite to the expected behaviour by RFM. 

\begin{figure}
 \includegraphics[width=\columnwidth]{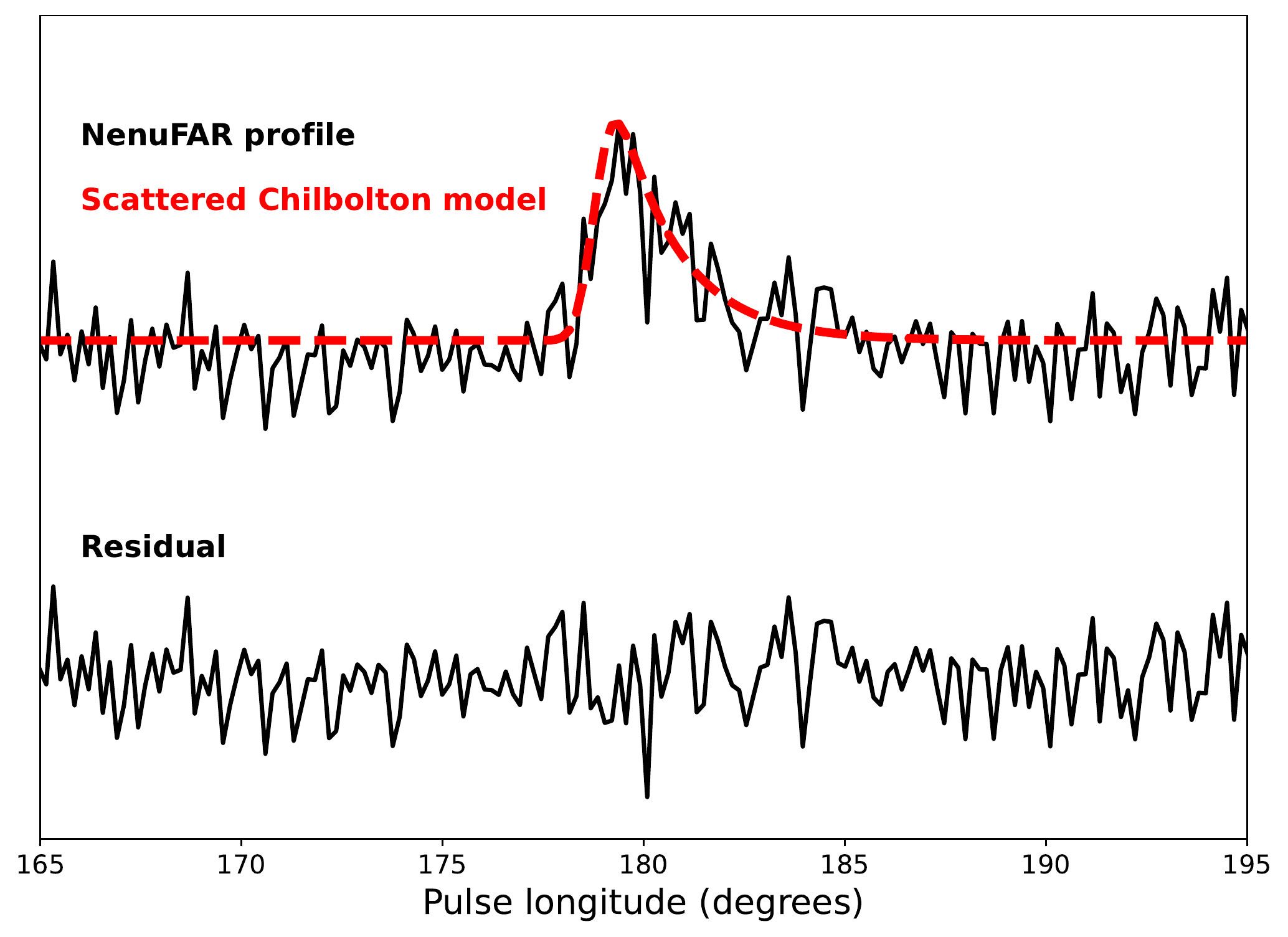}
 \caption{The NenuFAR profile (top) compared to a smoothed UK608 profile convolved with a scattering tail (dashed model curve, red in the online version). No significant signal remains in the residuals (bottom).}
 \label{fig: nenufar scattering}
\end{figure}

The NenuFAR profile, corresponding to the lowest frequency, is broader again and distinctly skewed. Given the steep rise followed by an exponentially decreasing tail, this can be attributed to scattering of the emission in the ISM. This is a strongly frequency-dependent effect with a power-law relationship between the scattering timescale and frequency, with a power law index of around $-4$ \citep{SDO1980, PulsarAstronomy,GK+2017}. This suggests that the scattering timescale for the NenuFAR data is around 50 times greater than at the UK608 centre frequency, which explains why only the NenuFAR profile is significantly affected. The NenuFAR profile is consistent with an intrinsic profile width which is equal to that observed at $\sim$150~MHz, albeit broadened by scattering. This is demonstrated in Fig.~\ref{fig: nenufar scattering} where the NenuFAR profile is compared with a von Mises function with a width equal to that of the UK608 profile, and convolved with an exponential scattering tail with an e-fold timescale of 0.1~s. This is consistent with the observed relationship between the DM and scattering timescale \citep[e.g.][]{BC+2004,IJW2019}. Therefore, scattering can fully explain the observed frequency evolution of the profile between 60 and 150~MHz (although given that its low signal-to-noise ratio makes it nearly impossible to resolve the profile reliably across the frequency band, the possibility of intrinsic profile evolution is not fully excluded). The high signal-to-noise (S/N) LOFAR Core profile also shows a somewhat elongated tail. No evolution of this tail is observed across the bandwidth of the observation, so we conclude that only the NenuFAR profile shows clear evidence for being scatter broadened.

\begin{figure}
 \includegraphics[width=\columnwidth]{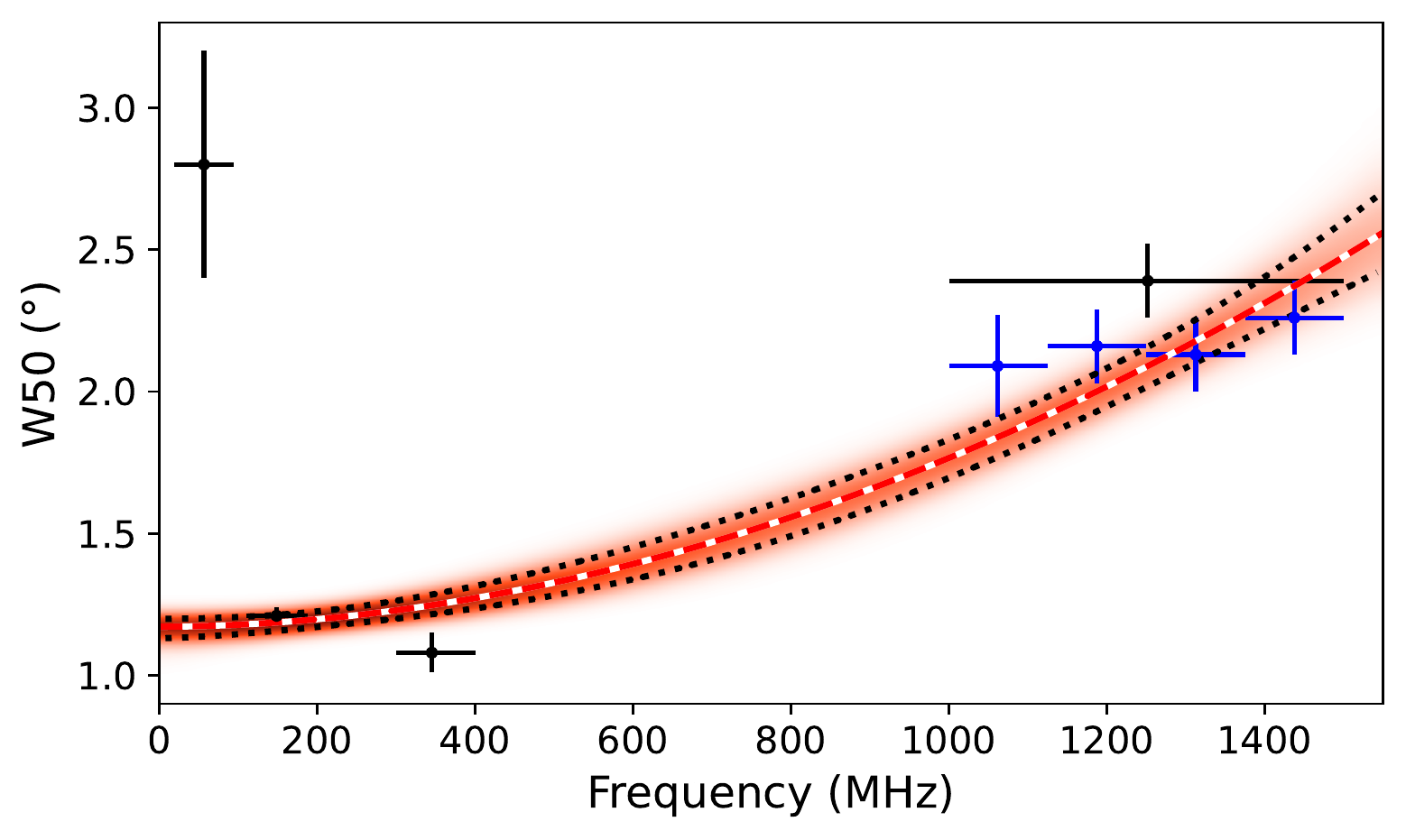}
 \caption{Evolution of the profile width of PSR~J0250+5854 with observing frequency. Points in black correspond to the profiles shown in Fig.~\ref{fig: profiles} and blue points are the profile widths of the four FAST sub-bands. The red and white dashed line represents the model of frequency evolution in Eq.~\eqref{eq: thorsett} which was fitted to the FAST sub-bands, GBT, UK608, and LOFAR Core data (NenuFAR was excluded because it is affected by scattering). A distribution of fits was calculated using bootstrapping techniques, and is represented by the red colour gradient. The black dotted lines indicate the 68~per~cent confidence interval of this distribution. The horizontal error bars indicate the bandwidth of a given observation.}
 \label{fig: width evolution}
\end{figure}

To quantify the pulse broadening at higher frequencies, we measured the profile widths as shown in Fig.~\ref{fig: profiles} by fitting von Mises functions to each profile using \textsc{psrsalsa}. This smooth mathematical description of the profile allows the width to be measured without being strongly affected by (white) noise. Two components were used to model the higher S/N profiles (LOFAR Core, GBT) but including more than one component for weaker profiles would result in over-fitting. The uncertainties on each measurement was calculated using bootstrapping where for each iteration a rotated version of the baseline was added to the profile. This ensures that both the statistical noise arising from the white noise as well as residual baseline variations are accounted for. The estimated full width at half maximum ($W_{50}$) of the profiles in Fig.~\ref{fig: profiles} are shown in Tab.~\ref{tab: W50}.
\begin{table}
\centering
\caption{The profile width at half maximum ($W_{50}$) of PSR~J0250+5854 as a function of frequency, measured by fitting von Mises functions to each profile shown in Fig.~\ref{fig: profiles}. These measurements are taken from the profiles as observed, and so includes the effect of scatter broadening in the case of the NenuFAR profile.}
\label{tab: W50}
\begin{tabular}{lcc}
    \hline
    Telescope & Centre freq. (MHz) & $W_{50}~(\degr)$ \\
    \hline
    FAST & 1250 & $2.4\pm0.1$ \\
    GBT & 350 & $1.08\pm0.07$ \\
    DE601 & 154 & $0.9\pm0.1$ \\
    UK608 & 150 & $0.9\pm0.1$ \\
    LOFAR Core & 149 & $1.21\pm0.03$ \\
    NenuFAR & 57 & $2.8\pm0.4$ \\ 
\end{tabular}
\end{table}
To further investigate the frequency evolution of the profile width, the widths were also determined after dividing the FAST data into four frequency sub-bands. Figure~\ref{fig: width evolution} shows the profile width against frequency for the profiles shown in Fig.~\ref{fig: profiles} (black) and the FAST sub-bands (blue).

The evolution of profile width with frequency $\nu$ was modelled using the relation
\begin{equation}
\label{eq: thorsett}
    W_{50} = A\nu^B + C,
\end{equation}
where $A$, $B$, and $C$ are constants \citep{Txx1991, CWx2014}. We fit the function in Eq.~\eqref{eq: thorsett} to the profile widths measured from the LOFAR Core, UK608, GBT, and FAST sub-band data. The NenuFAR profile was omitted to avoid scattering in the ISM affecting the results, and the DE601 and UK608 data were omitted because of their low S/N compared to the LOFAR Core data at a similar frequency. The horizontal error bars represent the bandwidth of a given observation. During fitting of Eq.~\eqref{eq: thorsett} the frequency of each observation was allowed to vary uniformly within these limits to account for the frequency-dependence of the profile width within the observed band. The distribution of fitted trend lines is shown in the red gradient plot in Fig.~\ref{fig: width evolution}. The black dotted lines bound the 68~per~cent confidence interval of the distribution, as a function of frequency. The red dashed line represents the optimal fit to the data, and the power-law exponent of Eq.~\eqref{eq: thorsett} is $1.9 \pm 0.4$. These findings are discussed further in Sec.~\ref{sec: discussion}.

\subsection{Polarisation and geometry}
\label{sec: polarisation and geometry}

The FAST data were calibrated using the pulsed noise diode signal. After calibration, the polarised FAST pulse profile of PSR~J0139+5814 (not shown, see also Sec.~\ref{sec: observations}) is in excellent agreement with the results of \citet{GLx1998} (publicly available on the European Pulsar Network (EPN) database\footnote{\url{http://www.epta.eu.org/epndb/}}). The LOFAR Core data were not polarisation-calibrated using the LOFAR station beam model, but rather using tied-array addition which incorporates data from different tiles and stations using the station calibration tables to account for the delays between them \citep[more detail can be found in ][]{SB+2019}. The signs of Stokes $V$ and the position angle curve had to be flipped in order to agree with convention \citep[e.g.][]{EWx2001}, a correction that was also applied to the FAST data. All data were corrected using the Faraday rotation measure $\mathrm{RM}=-54.65\pm0.02$~rad~m$^{-2}$, measured by applying RM synthesis \citep{BBx2005} to the LOFAR Core polarisation data. This is consistent with the RM measured using the FAST data, although this has a larger uncertainty due to the higher observing frequency.

\begin{figure}
 \includegraphics[width=\columnwidth]{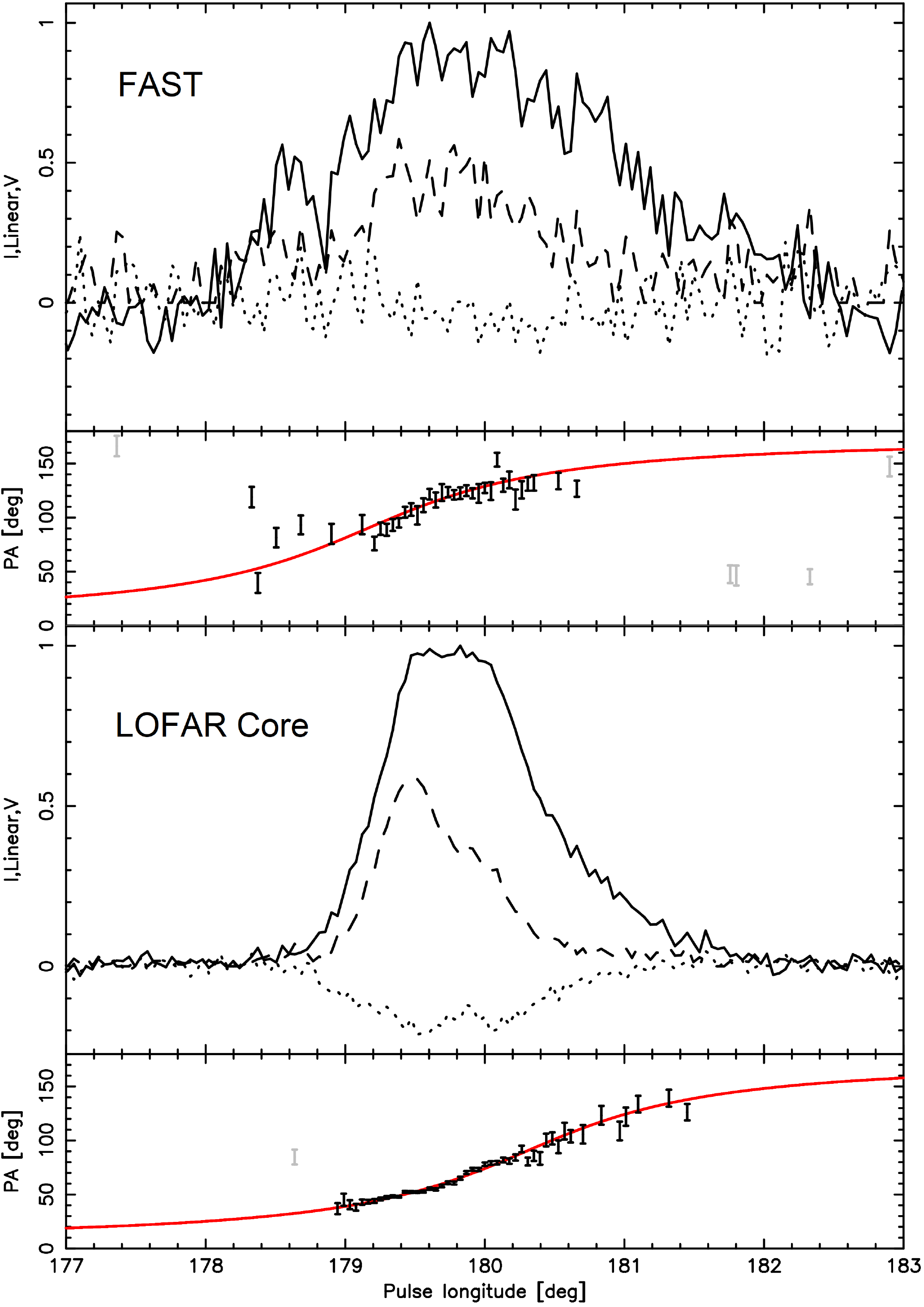}
 \caption{The polarised profile of PSR~J0250+5854, observed with FAST (upper plot) and LOFAR Core (lower plot). Total intensity is shown as the solid line, and linear and circular polarisation as dashed and dotted respectively. The lower panel of each plot shows the variation of position angle of linear polarisation with pulse longitude. The RVM (red curve in the online version) was fitted to the LOFAR Core data, and the same curve (with appropriate horizontal and vertical offset applied; see text) is shown for the FAST data. Although more are visible, only the darker PA points were used in fitting the RVM.}
 \label{fig: polarised profiles}
\end{figure}

In Figure~\ref{fig: polarised profiles} the polarised profile of PSR~J0250+5854 is shown as observed with FAST (first panel), and the LOFAR Core data (third panel). In both, the solid line is total intensity. The pulse profile has a moderate degree of linear polarisation (dashed), which was de-biased according to \citet{WKx1974}. There is negative circular polarisation (dotted line) in the LOFAR observation, and a hint of the same in the FAST data. The position angle (PA) $(\psi)$ as a function of pulse longitude is shown in the second and fourth panels of Fig.~\ref{fig: polarised profiles}, which relates to the Stokes $Q,\ U$ parameters via $\psi = 0.5 \arctan(U/Q)$. Its functional shape can be explained by the Rotating Vector Model \citep[RVM;][]{RCx1969}, a geometric model which links the observed changes in PA with pulse longitude $(\phi)$ to the orientation of the magnetic field lines with respect to the observer.

To fit the RVM, a grid search was conducted over the inclination angle of the magnetic axis, $\alpha$, and the impact parameter of the observer's line of sight with respect to the magnetic axis, $\beta$. For details, see Appendix~\ref{app: expanded geometry derivation} and \citet{RWJ2015a}. This was done for the LOFAR Core observation, and the best fit to the observed PA points is shown in Fig.~\ref{fig: polarised profiles} for both the LOFAR and FAST data after applying an offset in PA to account for the fact that no absolute PA calibration has been performed, and allowing for a shift of the inflection point in longitude. Only the darker PA points in Fig.~\ref{fig: polarised profiles} were used for fitting, as the PA points in the wings are uncertain and could be affected by orthogonal polarisation mode transitions \citep[e.g.][]{MSx2000}. The functional shapes of the LOFAR and FAST PA data are consistent, as expected when a dipolar field line configuration determines the shape. We therefore will only consider the RVM fit to the higher S/N LOFAR Core data.  As will be further discussed in Sec.~\ref{sec: discuss geometry}, an offset in the PA inflection point between the FAST and LOFAR Core data $\Delta\phi_0 = 1.1\pm0.8\degr$ was measured.  In Appendix~\ref{app: expanded geometry derivation} a more detailed analysis of the PA data can be found, and the main points are summarised here.

Given the very small duty-cycle, limited information is available about how the PA varies with pulse longitude, and as a consequence $\alpha$ and $\beta$ are highly correlated. The change of PA with pulse longitude is most rapid at the inflection point, $\sim$55~deg~deg$^{-1}$, which is predicted by the RVM to be equal to $\sin\alpha / \sin \beta$ \citep{Kxx1970}. This implies that $\beta$ must be small ($\lesssim1.8\degr$) as expected for a detection of a slowly rotating pulsar with a narrow beam directed along the direction of the magnetic axis. The magnetic inclination angle $\alpha$ is unconstrained from RVM fitting alone.

The measured profile widths provide additional constraints. Assuming the wider profile observed at FAST covers the full extend of the open field line region in a dipole geometry and the emission height lies within the range of 200 to 400~km \citep[e.g.][]{MRx2002, JKx2019}, then this corresponds to a half-opening angle of the beam $\rho \eqsim 1 - 2\degr$ (see Appendix~\ref{app: expanded geometry derivation}). This is consistent with what is expected for conal emission \citep{Rxx1993}, see Appendix~\ref{app: corecone model consistency}. This expectation for $\rho$ is related to the observed profile width via $\alpha$ and $\beta$, leading to the constraint $20 \lesssim \alpha \lesssim 50\degr$, or $130 \lesssim \alpha \lesssim 170\degr$. The frequency evolution of the beam will be discussed in Sec.~\ref{sec: discuss beam shape}.

\subsection{Flux density spectrum}
\label{sec: flux}

With a factor of $\sim$5 increase in spectral coverage with respect to \citet{TB+2018}, the radio spectrum of PSR~J0250+5854 could be further quantified. Flux calibration of the FAST data was possible by utilising the observation of the nearby BL Lacertae object J0303+472, which was used as a reference source (see Sec.~\ref{sec: observations}). This source, also known by the identifier 4C~47.08 \citep{VVx2006}, has a known flux density of 1.8~Jy at a wavelength of 20~cm (approximately 1500~MHz, suitably close to the centre frequency of the FAST data at 1250~MHz) as listed in the VLA Calibrator List\footnote{\url{https://science.nrao.edu/facilities/vla/observing/callist}}. The NASA/IPAC Extragalactic Database (NED)\footnote{\url{https://ned.ipac.caltech.edu/}} entry for this object contains a list of flux densities of this source at different frequencies from the literature. This reveals a significant scatter in flux density measurements of observations at similar frequencies. To accommodate this, as well as the uncertainty in the intrinsic flux because of interstellar scintillation, we assign an uncertainty of 50~per~cent to the flux density. This is consistent with other work on pulsar flux density measurements \citep[e.g.][]{Sxx1973}. The flux density calibration was performed using \textsc{psrchive}, and we measured a flux density of $4\pm2$~$\upmu$Jy for PSR~J0250+5854 at 1250~MHz. The S/N of the profile is also consistent with what is predicted for this flux density by the radiometer equation \citep[e.g.][]{handbook} with known (zenith angle dependent) values for the gain $G = 14$~K~Jy$^{-1}$ and system temperature $T_\mathrm{sys} = 25$~K of FAST\footnote{see also Appendix 2 of \url{http://english.nao.cas.cn/focus2015/201901/t20190130_205104.html}} \citep{LW+2018}. This flux density is below the upper limits at a similar frequency based on non-detections with the Lovell and Nan\c{c}ay telescopes \citep{TB+2018}.

\begin{figure}
 \includegraphics[width=\columnwidth]{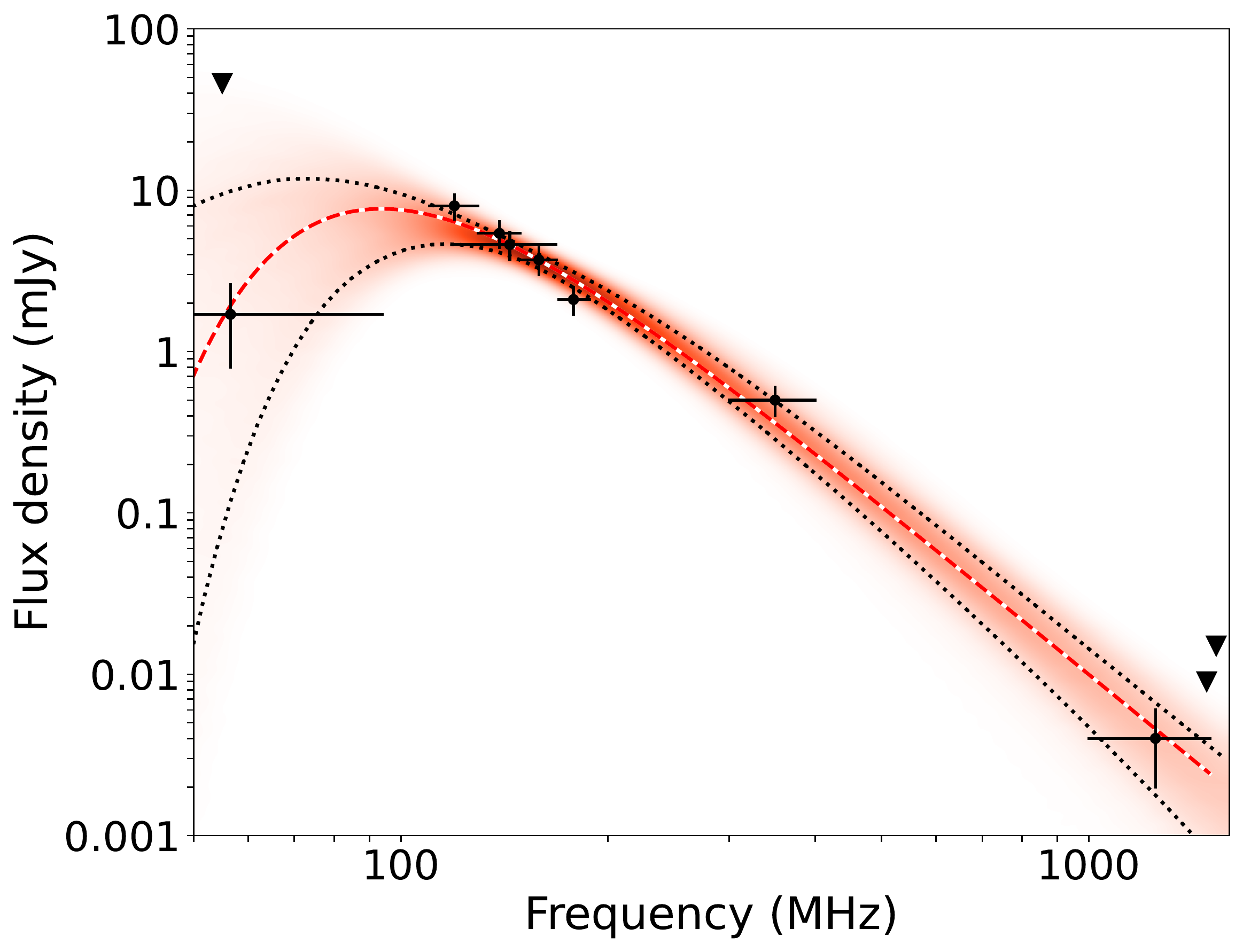}
 \caption{The flux density spectrum of PSR~J0250+5854, including upper limits (inverted triangles) from previous non-detections \citep{TB+2018}. As in Fig.~\ref{fig: width evolution}, the horizontal error bars indicate the bandwidth of a given observation. This plot includes the previous flux density measurements from Tan et al., with the new addition of the NenuFAR and FAST measurements. A power-law relationship with a low-frequency turnover was fitted to the data, and the best fit is indicated by the red and white dashed line. The distribution of acceptable fits is indicated, similar to Fig.~\ref{fig: width evolution}. The NenuFAR bandwidth extends down to 19~MHz, beyond what is shown in the figure. The fitted parameters from Eq.~\eqref{eq: spectrum} are the scaling factor $b = 0.1^{+0.3}_{-0.0}$, turnover parameter $m = 2.1^{+0.0}_{-1.2}$, critical frequency $\nu_c = 94\pm24$~MHz, and spectral index $k = -3.5^{+0.2}_{-1.4}$.}
 \label{fig: spectrum}
\end{figure}

Fig.~\ref{fig: spectrum} shows the flux density of PSR~J0250+5854 as a function of observing frequency, and includes the flux densities previously measured by \citet{TB+2018}. These previous measurements include detections with the GBT, LOFAR HBAs, and a flux density measurement obtained from the LOFAR Two-meter Sky Survey \citep[LoTSS;][]{SR+2017}. 
The detection of PSR~J0250+5854 at 57~MHz using NenuFAR marks the lowest frequency detection that is published. The flux density of the pulsar at this frequency was estimated using the radiometer equation, and was found to be $1.7\pm0.9$~mJy, where we have again assigned a 50~per~cent uncertainty. In calibrating these data the elevation of the source and number of antennas in the array were taken into account as they affect the gain, as does the bandpass of the array. The sky background temperature was estimated to be 9050~K at the position of PSR~J0250+5854 (which dominates over the receiver temperature of 776~K), found by extrapolating the sky temperature measured at 408~MHz \citep{HS+1982} with a spectral index of $-2.55$ \citep{LM+1987, RRx1988} to the centre frequency of 56.54~MHz. Full details of the NenuFAR flux calibration procedure are to be published in the instrumentation paper (Zarka et al., in prep.). This measurement indicates that the spectrum of PSR~J0250+5854 rolls over at low frequencies, and this explains the upper limit at a similar frequency reported by \citet{TB+2018} based on LOFAR Low Band Antenna array observations. We further discuss the spectral shape in Sec.~\eqref{sec: discuss flux}.

\section{Discussion}
\label{sec: discussion}

\subsection{Flux density and spectral index}
\label{sec: discuss flux}

\citet{TB+2018} were able to detect PSR~J0250+5854 over the frequency range between 120 and 350~MHz, and fitted a spectral index of $-2.6\pm0.5$, which is on the steeper side of the population distribution. However, with no detection of the pulsar at higher frequencies (1484~MHz, 1532~MHz) using the Lovell and Nan\c{c}ay radio telescopes (respectively), nor a detection with the core LOFAR Low Band Antenna stations (55~MHz), uncertainty remained over the broadband shape of the radio spectrum.

PSR~J0250+5854 is weak at a centre frequency of 57~MHz as observed with the NenuFAR telescope (compared to its flux density at 150~MHz), which is the result of a spectral turnover (see Sec.~\ref{sec: flux}). This is not unusual in the pulsar population: for example, \citet{BK+2016} noted that 25~per~cent of their low-frequency sample were fitted by a broken power law with a turnover typically around 100~MHz. Furthermore, \citet{JS+2018} noted that 21~per~cent of their sample deviates from a simple power law, exhibiting mainly broken power laws or low-frequency turnovers. The physical reasons for this are uncertain, but their analysis suggests that the deviations are partially intrinsic to the pulsar emission or because of magnetospheric absorption processes, and partially due to the environment around the pulsar or the ISM. 

To quantify the spectral turnover of PSR~J0250+5854, a power-law with a low-frequency turnover was fitted to the data using the same model used by \citet{JS+2018}, which is of the form
\begin{equation}
\label{eq: spectrum}
    S_\nu = b \bigg(\frac{\nu}{\nu_0}\bigg)^k \exp\bigg( \frac{k}{m} \bigg(\frac{\nu}{\nu_c}\bigg)^{-m}\bigg),
\end{equation}
where $\nu_0 = 500$~MHz is a constant (and arbitrary) reference frequency. The fitted parameters are $b$, a constant scaling factor; $k$, the spectral index; $\nu_c$, the turnover frequency; and $m$ which determines the smoothness of the transition. The value of $m$ is expected to be positive, and $\leq2.1$.

With only one flux density measurement below the turnover frequency, the parameters are somewhat ill-defined. The optimal fit\footnote{The resulting probability density function of $m$ is highly clustered at the maximum allowed value of $2.1$ which was implemented as a prior. Therefore, no meaningful uncertainty on $m$ could be assigned.} (Fig.~\ref{fig: spectrum}, red line) is for an exponent $m=2.1$. This corresponds to the sharpest turnover allowed within the free-free absorption model (see \citet{JS+2018} and references therein). The fitted spectral index of $k = -3.5^{+0.2}_{-1.4}$ is steep compared to the mean found for the pulsar population (\citealt{BLV2013} found a mean spectral index of $-$1.4 with unit standard deviation, whilst \citealt{JS+2018} found a mean of $-$1.60 with a standard deviation of 0.54). But other examples of such steep spectral indices exist, including PSR~J1234$-$6423 which has a broken power-law spectrum with a spectral index of $-3.8\pm0.5$ below $\sim$1700~MHz \citep{JS+2018}.

\citet{TB+2018} noted that there are occasional bright pulses at 350~MHz in the leading component of the profile of PSR~J0250+5854. This behaviour was not seen at around 150~MHz. If equally bright single pulses exist in the FAST frequency band, then they should be comfortably detectable above the level of the thermal noise. However, the residual baseline variations in the single pulse data are such that these bright pulses cannot be confidently detected. It therefore remains to be seen how the erratic nature of the single pulses evolves above frequencies of 350~MHz.


\subsection{Profile width evolution}
\label{sec: discuss width} 
As can be seen in Fig.~\ref{fig: width evolution}, the profile width of PSR~J0250+5854 increases with observing frequency, from around $1\degr$ at 150~MHz to $2\degr$ at 1250~MHz. Eq.~\eqref{eq: thorsett} was fitted to the measured profile widths as a function of frequency, resulting in a power-law index $B = 1.9\pm0.4$ (see Sec.~\ref{sec: profile widths}). Although it is unusual for pulsars to have a positive index, meaning that their profiles broaden with increasing frequency, there are other examples. \citet{CWx2014} identified 29 pulsars out of 150 with such a positive index based on profiles from the EPN database. Given the relatively large uncertainty on both our measured value of $B$ and those measured by \citet{CWx2014}, the index for PSR~J0250+5854 is consistent with 24 out of the 29 pulsars with a reported positive index. None of these 29 pulsars have a significantly larger $B$ than that of PSR~J0250+5854. This increase in profile width with frequency is contrary to the expectation from RFM.

Higher frequency radiation can be expected to be produced closer to the neutron star. This follows from models based on curvature radiation from relativistic bunches of particles travelling along the magnetic field lines \citep[e.g.][ and references therein]{GLM2004,DRx2015}, as well as those based on plasma instabilities since both the plasma density and the plasma frequency decrease with increasing altitude (e.g. \citealt{HAx2001} and references therein; also \citealt{GGM2002}). As a consequence, the opening angle of the radio beam can be expected to be narrower at higher frequencies, unless a larger fraction of the open-field-line region becomes active. \citet{PH+2016} studied 100 pulsars and measured their profile widths at frequencies ranging from tens of megahertz up to 1400~MHz. Only in a few cases was the profile width seen to increase significantly with frequency. In those pulsars, profile broadening was indeed caused by the emergence of new profile components as frequency increased. 

Observations at a frequency around 800~MHz could help reveal the reasons for the abnormal frequency evolution of the profile of PSR~J0250+5854. The profile morphology of PSR~J0250+5854 is indeed complex, with a profile shape skewed in both the LOFAR Core and FAST observations. The GBT profile shows a distinct double-peaked structure with distinct behaviours since \citet{TB+2018} noted that the stronger first component was caused by occasional strong individual pulses. At other frequencies no well separated profile components are observed. However, the flattened peak in the LOFAR Core profile is suggestive of two blended components of similar intensity. This flattening was not visible for the profiles published in \citet{TB+2018}, which is because of the lower S/N. By inspecting all available data, no significant profile shape variability has been detected in LOFAR Core observations of PSR~J0250+5854.

\subsection{Emission height}
\label{sec: discuss geometry}
Constraining the viewing geometry is particularly interesting for slowly rotating pulsars to highlight differences with magnetars (see Sec.~\ref{sec: discuss compare}). As seen in Fig.~\ref{fig: polarised profiles}, the inflection point of the PA swing occurs close to the centre of both the FAST and LOFAR profiles. The lack of significant relativistic aberration and retardation (A/R) shift \citep{BCW1991} which moves the inflection point towards, or even beyond, the edge of the profile implies that the emission height must be considerably smaller than the light cylinder radius. This affirms that radio pulsars produce emission at an absolute emission altitude that is relatively constant across the population, rather than being at a constant fraction of $R_\mathrm{LC}$. Given the emission height is not at a constant fraction of $R_\mathrm{LC}$ there should be a period dependence of the pulse width \citep[e.g.][]{Rxx1993}. Based on this relationship \citet{KJx2007} proposed that the maximum emission height of radio pulsars at 1.4~GHz is around 1000~km, refined to an absolute height range of 200 to 400~km irrespective of pulse period \citep{JKx2019, JSK2020} -- this is the range we assume for the FAST profile (1250~MHz) in Sec.~\ref{sec: polarisation and geometry}.

There is an indication of a difference in the longitude of the inflection point of the PA curve between the LOFAR and FAST frequencies of $\Delta \phi_0 = 1.1\pm0.8\degr$. The direction of the shift suggests that lower frequencies are produced higher in the magnetosphere in line with conventional RFM, which would make the narrowness of the low frequency profiles even more striking. A higher S/N detection at the FAST frequency could make this measurement more significant, and clarify if there are systematic inconsistencies in the PA swing compared to the prediction of the RVM which could play a role.

\subsection{Beam shape evolution and polar cap configuration}
\label{sec: discuss beam shape}

The profile width evolution of PSR~J0250+5854 with frequency is strong, and opposite compared to what is expected from conventional RFM. As pointed out in Sec.~\ref{sec: discuss geometry}, such behaviour is often linked to the emergence of new profile components as frequency increases. Although for PSR~J0250+5854 there is no strong evidence of the emergence of a new profile component, it is clear the profile morphology is complex (Sec.~\ref{sec: profile widths}). Under the assumption that emission height increases with decreasing frequency, the implication is that at lower frequencies the beam is underfilled compared to the open field line region. This means that only a small fraction of the field lines are significantly active, and emission is more concentrated to a specific region of the polar cap. This could be associated with a frequency dependence in the active patches in a patchy beam model \citep[e.g.][]{LMx1988, KSx2007}. Here we outline other alternative models of the beam structure that could explain our observations.

In the framework of the core-cone model \citep[e.g.][]{Rxx1983a,Rxx1983b, RRx1990, Rxx1993} the broadening of the pulse profile with increasing frequency can be explained by so-called conal outriders. Here, the more narrow central core component dominates at low frequencies. However, the wider conal components with a more shallow spectrum become more prominent at higher frequencies. The rise of conal outriders at FAST frequencies then must be blended with the core component to explain the wide single-peaked profile.
In this scenario it is expected that the profile would evolve into a wide, double-peaked profile at even higher radio frequencies. The bifurcated GBT profile in Fig.~\ref{fig: profiles} would therefore not be a consequence of conal emission, but should instead be associated with a magnetospheric absorption feature of a core-single profile, which are known to occur between 200 to 800~MHz \citep[e.g.][]{Rxx1983b, Rxx1986}. 
In Appendix~\ref{app: corecone model consistency} the viewing geometry is derived in this framework, leading to results consistent with Sec.~\ref{sec: polarisation and geometry} and Appendix~\ref{app: expanded geometry derivation}.
Unlike most pulsars with conal outriders \citep[e.g.][]{RSW1989}, PSR~J0250+5854 has a low energy loss rate, $\dot{E}\sim 10^{29}$~erg~s$^{-1}$ (compared to $\dot{E} \gtrsim 10^{32}$~erg~s$^{-1}$).

A non-uniform emission height could amplify the effects of a frequency dependence in emissivity across the open field line region. There is evidence for the emission height to be larger near the rim of the open field line region \citep[e.g.][and references therein]{GGx2003, WJx2008b}. More recently, \citet{ROWx2020} linked the core-cone model to the plasma pair multiplicity model of \citet{THxx2015} in which the pair production front can form a cup-like structure in the open field line region. The increase in emission height towards the rim of the open field line region further increases the beaming fraction of emission. With the suggestion that the emission height for PSR~J0250+5854 at LOFAR frequencies is higher compared to FAST frequencies (see Sec.~\ref{sec: discuss geometry}) this may not be a significant effect.

\citet{CWx2014} conclude that in a number of pulsars the widening of the profile at higher frequencies could not be ascribed to structures consistent with the core-cone model. This also appears to be the case in a small sub-group of pulsars studied by \citet{PH+2016}. Like PSR~J0250+5854, these pulsars do not show well-separated profile peaks at the highest frequencies. The expectation from conventional RFM is based on emission being produced over a wide range of altitudes, with each height producing narrowband emission. On the other hand, if a narrow range of emission heights generates broadband emission the observed spectrum will be different. This led \citet{CWx2014} to suggest that fan beams could accommodate anti-RFM-like behaviour. Broadband emission is incorporated into the fan beam model \citep{Mxx1987, DRD2010, DRx2012, DRx2013, WP+2014} where emission is produced along magnetic flux tubes that extend out from the pole in a fan-like structure. A single fan, hence spark, would be required to explain the beam structure of PSR~J0250+5854. Support for this beam structure is found in observations of the precessing pulsars J1141$-$6545 and J1906+0746 \citep{MK+2010, DK+2013}. Following the suggestion of \citet{Mxx1987} that each flux tube may have its own spectrum, \citet{CW+2007} argued that the emission spectrum may not be homogeneous across a flux tube. In particular, they argue that pulsars which show pulse broadening with increasing frequency may have a flattening emission spectrum away from the magnetic axis, as supported by their simulations. Broadband emission in flux tubes with a location-varying spectral index follows naturally from particle-in-cell simulations of vacuum-gap pair-production \citep{Txx2010} which predict that the momentum spectrum of the secondary plasma is not necessarily monotonic as a function of height within the magnetosphere, and so a given observed frequency cannot be assigned to a unique altitude.

As highlighted in the introduction, the slow rotation of PSR~J0250+5854 implies a tiny polar cap connected to the open field line region. One can wonder if there is enough space to fully develop the type of beam complexity encountered in more typical pulsars. In the model of \citet{MB+2020} it is predicted that the polar cap of PSR~J2144$-$3933 is only large enough to support a single pair-production site -- known as a ``spark'' -- hence its profile would be a single component. For PSR~J0250+5854 (which although slower has a much larger $\dot{P}$) their model predicts that the footprint of a spark is one fifth of the area of the polar cap, meaning that up to three sparks could be supported if they are packed tightly. 
Moreover, sparks are also required to be separated from one another due to screening effects \citep[e.g.][]{GSx2000}, meaning that the polar cap may only be able to support a single spark. This single spark as proposed for PSR~J2144$-$3933, may not circulate about the magnetic axis in a way theorised for multi-spark systems \citep{RSx1975}. Such a single-spark scenario may explain why while the profile width evolves as a function of frequency, the profile essentially remains single peaked.

\subsection{Comparison to other slow pulsars and magnetars}
\label{sec: discuss compare}

The extremely long period of PSR~J0250+5854 places it on the far right-hand side of the $P$-$\dot{P}$ diagram, in an area largely inhabited by magnetars and X-ray Dim Isolated Neutron Stars (XDINSs). These objects are detected only as soft thermal X-ray sources without radio counterparts. Of the seven brightest XDINSs, five have high magnetic dipole fields of the order of $10^{13}$--$10^{14}$~G which may mean they are related to magnetars \citep{Hxx2007,KKx2007}. Despite the fact that magnetars form a distinct class of objects with much greater spin-down rates, and hence much higher rates of loss of rotational energy $\dot{E}$, they may evolve with time towards the parameter space occupied by the slow pulsars \citep[e.g.][]{VR+2013}. However, to date PSR~J0250+5854 remains undetected in X-rays, despite a dedicated \textit{Swift} X-Ray Telescope observation, which makes it difficult to confirm a connection between it and XDINS (see \citealt{TB+2018} for details). Similarly, PSR~J0250+5854 has not shown any magnetar-like behaviour such as bursts, or large radio variability as of yet. Furthermore, PSR~J0250+5854 is located squarely in the Galactic plane which argues that it is still relatively young.

Nevertheless, since the pulse period is a key factor controlling the width of radio pulse profiles, it is worth comparing PSR~J0250+5854 with the radio-emitting magnetars and other slowly spinning rotation-powered radio pulsars. This comparison can highlight what other parameters play a role in the radio beam geometry of these slowly rotating objects. Aside from PSR~J0250+5854, the two other slowest-spinning known radio pulsars are PSRs~J2251$-$3711 ($P=12.1$~s), and J2144$-$3933 ($P=8.5$~s). There are five known magnetars for which pulsed radio emission has been detected. A summary of their properties is shown in Table~\ref{tab: comparison}.
\begin{table*}
	\centering
	\caption{Parameters (period $P$, spin-down rate $\dot{P}$) of the three slowest-spinning radio pulsars (top three sources) and the five magnetars (lower five sources) known to produce pulsed radio emission. The profile width of the radio pulsars are measured values of $W_{50}$ taken from this work and the referenced literature. The widths of the magnetar profiles are estimated based on the full extent over which significant emission was visible in the published profiles (at the reference frequency) in order to capture the complexity of the magnetar profiles. \newline \textbf{References:} (1) \citet{YMJ1999}; (2) \citet{MB+2020}; (3) \citet{MK+2020}; (4) \citet{CR+2007a}; (5) \citet{CR+2008}; (6) \citet{LB+2010}; (7) \citet{LB+2012}; (8) \citet{EF+2013}; (9) \citet{CR+2006}; (10) \citet{CR+2007b}; (11) \citet{KS+2007}; (12) \citet{LL+2019}; (13) \citet{ER+2020}; (14) \citet{LS+2020}; (15) \citet{CC+2020}.}
	\label{tab: comparison}
	\begin{tabular}{lrrrrr} 
		\hline
	    Object & $P$ (s) & $\dot{P}$ (ss$^{-1}$) & Profile Width ($\degr$) & Ref. Freq. (MHz) & References\\
		\hline
		PSR~J0250+5854          & 23.5 & $2.72\times10^{-14}$ &  $2.4$ & 1250 & This work\\
		PSR~J2144$-$3933        & 8.5  & $4.96\times10^{-16}$ &  $0.8$ & 1400 & 1, 2\\
		PSR~J2251$-$3711        & 12.1 & $1.31\times10^{-14}$ & $1.2$  & 1382 & 3\\
		\hline
		1E~1547.0$-$5408        & 2.1  & $2.32\times10^{-11}$ & $90$   & 6600 & 4, 5\\
        PSR~J1622$-$4950        & 4.3  & $1.70\times10^{-11}$ & $190$  & 1400 & 6, 7\\
        PSR~J1745$-$2900        & 3.8  & $6.80\times10^{-12}$ & $15$   & 2400 & 8\\
        XTE~J1810$-$197          & 5.5  & $1.02\times10^{-11}$ & $35$   & 1400 & 9, 10, 11, 12\\
        Swift~J1818.0$-$1607    & 1.4  & $9\times10^{-11}$    & $20$   & 1548 & 13, 14, 15\\
		\hline
	\end{tabular}
\end{table*} 
Although the focus here will be the differences in profile widths, there are other differences such as the spectra of magnetars being radically different \citep{CR+2007b, CR+2007a, KS+2007, KJ+2011,LS+2020, CC+2020}, and their radio emission being much more transient with periods of activity and strongly changing profile shapes \citep{SS+2009, DJ+2018, LL+2019, DL+2019,LJS+2021}.

The three slowest-spinning radio pulsars have long periods and narrow, fairly simple pulse profiles. Measurements of $W_{50}$ are published for these profiles and are representative of the overall profile width. Despite being the slowest-spinning of the three, PSR~J0250+5854 has the widest profile by a factor of around two. If only the period determines the width, the inverse would be expected. This suggests that besides PSR~J0250+5854 (see Sec.~\ref{sec: discuss geometry}) underfilling of the beam may also play a role in the other slow pulsars. In contrast, the magnetars have more complex profiles with distinct components, which means that $W_{50}$ is not always representative of the overall profile width. Therefore, for the magnetars in Table~\ref{tab: comparison} the full span (rounded to the nearest five degrees) over which emission is seen in the published profiles are reported.

There is a stark contrast between the profile widths of the magnetars compared to the slow pulsars, much more than can be expected from just the differences in $P$ (and hence $R_\mathrm{LC}$). There are three potential geometric explanations: 1) for the slowly rotating pulsars only a tiny fraction of the open-field-line region is active; 2) all magnetars have a magnetic axis almost aligned with the rotation axis; 3) all magnetars have atypically wide beams due to large emission heights or otherwise. We will argue that only options 2) and 3) are viable, and that option 4) plays a more significant role. We will not consider the very unlikely coincidence that all three slowly rotating pulsars are observed with extremely grazing lines of sight with respect to the radio beam. 

The active fraction of the open field line region needs to be very small for the slow pulsars if it is to be the main reason why the slow pulsars have such narrow beams compared to the magnetars. This seems unlikely, as it would not explain why there are no slow pulsars with multiple narrow profile components spread over a similar fraction of the rotation period for which magnetars show emission. The complex magnetar profiles often exhibit individual components which are much wider than the full profiles of the slow pulsars.

If magnetars have very aligned radio beams with respect to their rotation axis (small $\alpha$), the observer's line of sight would spend a larger fraction of the time within the beam. The process of Sec.~\ref{sec: polarisation and geometry} (detailed in Appendix~\ref{app: expanded geometry derivation}) can be used to estimate how extreme the alignment of the magnetars should be in order to explain their wide pulse profiles. Taking the mean magnetar period and median profile width from Tab.~\ref{tab: comparison} with an emission height of 400~km, such an object requires $\alpha\lesssim 11\degr$ to produce profiles of the observed width. However, there is little evidence for this. Polarisation studies have given a variety of $\alpha$ values for the magnetars ranging from near aligned to almost orthogonal \citep{CC+2007,KS+2007,CR+2008,LB+2012,LJS+2021}. In addition, highly aligned magnetars are difficult to reconcile with the large modulation of the thermal X-rays, as was highlighted for XTE~J1810$-$197 \citep{GHx2007,PGx2008} and 1E~1547.0$-$5408 \citep{IE+2010}. Furthermore, if magnetars evolve into slowly spinning rotation-powered pulsars no difference in their $\alpha$ distribution can be expected. For PSR~J0250+5854, for example, we have shown that it is unlikely that $\alpha$ is very small. Therefore, it is concluded that the emission height must play a significant role in explaining the magnetar radio profile widths.

If the emission heights are the dominant reason for the magnetars having wider radio profiles, they need to be $\sim$20 times larger (around 10,000~km) compared to the slowly spinning rotation-powered pulsars. Alternatively, if their radio beams are confined by last open field lines which close within the light cylinder (as suggested by detailed simulations by \citealt{Sxx2006}; see also the discussion of `Y-points' by \citealt{Cxx2014}), or if magnetic field sweepback plays a large role \citep[e.g.][]{CRx2012} then their polar beams will be wider as well. Therefore, one cannot distinguish between large emission heights and extended open field line regions \citep[e.g.][]{RWJ2015b, RWJ2015a}. In such a scenario, the open field line regions of magnetars would need to be $\sim$5 times larger than predicted for a static dipole field. In either case, it would imply that slowly spinning rotation-powered pulsars have dramatically reduced beaming fractions compared to magnetars which therefore plays an important role in the deficit of observed slow pulsars near the death valley.

The large implied difference in the beaming fraction between the two classes of objects is unexpected given the weak $\dot{P}$ dependence of the pulse widths observed for the normal pulsar population \citep[e.g.][]{KGx2003, JKx2019}. This implies that for these slowly rotating objects the role of $\dot{P}$ in governing the beaming fraction is much larger than for the normal pulsar population. This could potentially be facilitated by the incredible strengths of the magnetar magnetic fields.

\section{Conclusions}
\label{sec: conclusions}

We have obtained the highest- and lowest-frequency radio detections of PSR~J0250+5854, the most slowly rotating radio-emitting pulsar known, using simultaneous observations from 57~MHz to 1250~MHz. The highest frequency detection with FAST (1250~MHz) shows that the spectrum is exceptionally steep with a spectral index of $-3.5^{+0.2}_{-1.4}$ and the lowest frequency detection with NenuFAR (57~MHz) reveals a spectral turn-over below 95~MHz. While PSR~J0250+5854 is slow, it is relatively close to the `death valley' and the fact it is still active in radio can be accommodated in current emission models. The pulse profile shows narrowing at lower frequencies, contrary to the expectations of radius-to-frequency mapping. This decrease in beaming fraction is suggestive of a reduction in the filling fraction of the beam, or disappearing conal outriders. The polarisation information of LOFAR Core data at 150~MHz and FAST data at 1250~MHz was used to show that the line-of-sight impact parameter $\beta$ is very small, passing within $1.8\degr$ of the magnetic axis. This confirms that the radio beam is very narrow, as expected for such a slow pulsar. Furthermore, the lack of a delay between the profile peak and position angle curve inflection point implies that the emission height of PSR~J0250+5854 at 1250~MHz is low, consistent with those found for other non-recycled pulsars. Finally, we draw comparisons between other slow pulsars, PSR~J0250+5854, and the five known magnetars with pulsed radio emission which have the most similar pulse periods in the known pulsar population. To explain the significantly broader magnetar profiles magnetic alignment may play a role, but the main reason is likely to be either considerably expanded open field line regions or substantially larger emission heights for magnetars.

\section*{Acknowledgements}

This research has made use of the NASA/IPAC Extragalactic Database (NED) which is operated by the Jet Propulsion Laboratory, California Institute of Technology, under contract with the National Aeronautics and Space Administration. Pulsar research at Jodrell Bank Centre for Astrophysics and Jodrell Bank Observatory is supported by a consolidated grant from the UK Science and Technology Facilities Council (STFC). This research is also supported by the National Natural Science Foundation of China NSFC (Grant No. 11988101, No. U1938117, No. U1731238 and No. 11703003).

This paper is based on data from the German LOng-Wavelength (GLOW) array, which is part of the International LOFAR Telescope (ILT) which is designed and built by ASTRON \citep{lofarPaper}. Specifically, we used the Effelsberg (DE601) station funded by the Max-Planck-Gesellschaft. The observations of the German LOFAR stations were carried out in the stand-alone GLOW mode which is technically operated and supported by the Max-Planck-Institut f\"ur Radioastronomie, the Forschungszentrum J\"ulich, Bielefeld University, by BMBF Verbundforschung project D-LOFAR III (grant number 05A14PBA) and by the states of Nordrhein-Westfalen and Hamburg.

This paper is based on data obtained using the NenuFAR radio-telescope. The development of NenuFAR has been supported by personnel and funding from: Station de Radioastronomie de Nan\c{c}ay, CNRS-INSU, Observatoire de Paris-PSL, Universit\'e d'Orl\'eans, Observatoire des Sciences de l'Univers en r\'egion Centre, R\'egion Centre-Val de Loire, DIM-ACAV and DIM-ACAV+ of R\'egion Ile de France, Agence Nationale de la Recherche. We acknowledge the use of the Nan\c{c}ay Data Center computing facility (CDN - Centre de Donn\'ees de Nan\c{c}ay). The CDN is hosted by the Station de Radioastronomie de Nan\c{c}ay in partnership with Observatoire de Paris, Universit\'e d'Orl\'eans, OSUC and the CNRS. The CDN is supported by the Region Centre Val de Loire, d\'epartement du Cher. The Nan\c{c}ay Radio Observatory is operated by the Paris Observatory, associated with the French Centre National de la Recherche Scientifique (CNRS).
 
This paper is based on data obtained with the International LOFAR Telescope (ILT) under project code DDT8\_004. LOFAR \citep{lofarPaper} is the Low Frequency Array designed and constructed by ASTRON. It has observing, data processing, and data storage facilities in several countries, that are owned by various parties (each with their own funding sources), and that are collectively operated by the ILT foundation under a joint scientific policy. The ILT resources have benefited from the following recent major funding sources: CNRS-INSU, Observatoire de Paris and Universit\'{e} d'Orl\'{e}ans, France; BMBF, MIWF-NRW, MPG, Germany; Science Foundation Ireland (SFI), Department of Business, Enterprise and Innovation (DBEI), Ireland; NWO, The Netherlands; The Science and Technology Facilities Council, UK.

J.W.T.H. acknowledges funding from an NWO Vici grant (``AstroFlash'').

\section*{Data Availability}
The data underlying this article will be shared on reasonable request to the corresponding author.
\bibliographystyle{mnras}
\bibliography{0250_bibliography} 

\newcommand{\noop}[1]{}
\begin{thebibliography}{}
\makeatletter
\relax
\def\mn@urlcharsother{\let\do\@makeother \do\$\do\&\do\#\do\^\do\_\do\%\do\~}
\def\mn@doi{\begingroup\mn@urlcharsother \@ifnextchar [ {\mn@doi@}
  {\mn@doi@[]}}
\def\mn@doi@[#1]#2{\def\@tempa{#1}\ifx\@tempa\@empty \href
  {http://dx.doi.org/#2} {doi:#2}\else \href {http://dx.doi.org/#2} {#1}\fi
  \endgroup}
\def\mn@eprint#1#2{\mn@eprint@#1:#2::\@nil}
\def\mn@eprint@arXiv#1{\href {http://arxiv.org/abs/#1} {{\tt arXiv:#1}}}
\def\mn@eprint@dblp#1{\href {http://dblp.uni-trier.de/rec/bibtex/#1.xml}
  {dblp:#1}}
\def\mn@eprint@#1:#2:#3:#4\@nil{\def\@tempa {#1}\def\@tempb {#2}\def\@tempc
  {#3}\ifx \@tempc \@empty \let \@tempc \@tempb \let \@tempb \@tempa \fi \ifx
  \@tempb \@empty \def\@tempb {arXiv}\fi \@ifundefined
  {mn@eprint@\@tempb}{\@tempb:\@tempc}{\expandafter \expandafter \csname
  mn@eprint@\@tempb\endcsname \expandafter{\@tempc}}}

\bibitem[\protect\citeauthoryear{{Bates}, {Lorimer}  \& {Verbiest}}{{Bates}
  et~al.}{2013}]{BLV2013}
{Bates} S.~D.,  {Lorimer} D.~R.,   {Verbiest} J.~P.~W.,  2013, \mn@doi [\mnras]
  {10.1093/mnras/stt257}, \href
  {https://ui.adsabs.harvard.edu/abs/2013MNRAS.431.1352B} {431, 1352}

\bibitem[\protect\citeauthoryear{{Bhat}, {Cordes}, {Camilo}, {Nice}  \&
  {Lorimer}}{{Bhat} et~al.}{2004}]{BC+2004}
{Bhat} N.~D.~R.,  {Cordes} J.~M.,  {Camilo} F.,  {Nice} D.~J.,   {Lorimer}
  D.~R.,  2004, \mn@doi [\apj] {10.1086/382680}, \href
  {https://ui.adsabs.harvard.edu/abs/2004ApJ...605..759B} {605, 759}

\bibitem[\protect\citeauthoryear{{Bilous} et~al.,}{{Bilous}
  et~al.}{2016}]{BK+2016}
{Bilous} A.~V.,  et~al., 2016, \mn@doi [\aap] {10.1051/0004-6361/201527702},
  \href {https://ui.adsabs.harvard.edu/abs/2016A&A...591A.134B} {591, A134}

\bibitem[\protect\citeauthoryear{{Blaskiewicz}, {Cordes}  \&
  {Wasserman}}{{Blaskiewicz} et~al.}{1991}]{BCW1991}
{Blaskiewicz} M.,  {Cordes} J.~M.,   {Wasserman} I.,  1991, \mn@doi [\apj]
  {10.1086/169850}, \href
  {https://ui.adsabs.harvard.edu/abs/1991ApJ...370..643B} {370, 643}

\bibitem[\protect\citeauthoryear{{Bondonneau} et~al.,}{{Bondonneau}
  et~al.}{2020}]{LUPPI}
{Bondonneau} L.,  et~al., 2020, A\&A, in press.

\bibitem[\protect\citeauthoryear{{Brentjens} \& {de Bruyn}}{{Brentjens} \& {de
  Bruyn}}{2005}]{BBx2005}
{Brentjens} M.~A.,  {de Bruyn} A.~G.,  2005, \mn@doi [\aap]
  {10.1051/0004-6361:20052990}, \href
  {https://ui.adsabs.harvard.edu/abs/2005A&A...441.1217B} {441, 1217}

\bibitem[\protect\citeauthoryear{{Camilo}, {Ransom}, {Halpern}, {Reynolds},
  {Helfand}, {Zimmerman}  \& {Sarkissian}}{{Camilo} et~al.}{2006}]{CR+2006}
{Camilo} F.,  {Ransom} S.~M.,  {Halpern} J.~P.,  {Reynolds} J.,  {Helfand}
  D.~J.,  {Zimmerman} N.,   {Sarkissian} J.,  2006, \mn@doi [\nat]
  {10.1038/nature04986}, \href
  {https://ui.adsabs.harvard.edu/abs/2006Natur.442..892C} {442, 892}

\bibitem[\protect\citeauthoryear{{Camilo}, {Reynolds}, {Johnston}, {Halpern},
  {Ransom}  \& {van Straten}}{{Camilo} et~al.}{2007a}]{CR+2007b}
{Camilo} F.,  {Reynolds} J.,  {Johnston} S.,  {Halpern} J.~P.,  {Ransom} S.~M.,
    {van Straten} W.,  2007a, \mn@doi [\apjl] {10.1086/516630}, \href
  {https://ui.adsabs.harvard.edu/abs/2007ApJ...659L..37C} {659, L37}

\bibitem[\protect\citeauthoryear{{Camilo} et~al.,}{{Camilo}
  et~al.}{2007b}]{CC+2007}
{Camilo} F.,  et~al., 2007b, \mn@doi [\apj] {10.1086/518226}, \href
  {https://ui.adsabs.harvard.edu/abs/2007ApJ...663..497C} {663, 497}

\bibitem[\protect\citeauthoryear{{Camilo}, {Ransom}, {Halpern}  \&
  {Reynolds}}{{Camilo} et~al.}{2007c}]{CR+2007a}
{Camilo} F.,  {Ransom} S.~M.,  {Halpern} J.~P.,   {Reynolds} J.,  2007c,
  \mn@doi [\apjl] {10.1086/521826}, \href
  {https://ui.adsabs.harvard.edu/abs/2007ApJ...666L..93C} {666, L93}

\bibitem[\protect\citeauthoryear{{Camilo}, {Reynolds}, {Johnston}, {Halpern}
  \& {Ransom}}{{Camilo} et~al.}{2008}]{CR+2008}
{Camilo} F.,  {Reynolds} J.,  {Johnston} S.,  {Halpern} J.~P.,   {Ransom}
  S.~M.,  2008, \mn@doi [\apj] {10.1086/587054}, \href
  {https://ui.adsabs.harvard.edu/abs/2008ApJ...679..681C} {679, 681}

\bibitem[\protect\citeauthoryear{{Champion} et~al.,}{{Champion}
  et~al.}{2020}]{CC+2020}
{Champion} D.,  et~al., 2020, \mn@doi [\mnras] {10.1093/mnras/staa2764}, \href
  {https://ui.adsabs.harvard.edu/abs/2020MNRAS.498.6044C} {498, 6044}

\bibitem[\protect\citeauthoryear{{Chen} \& {Ruderman}}{{Chen} \&
  {Ruderman}}{1993}]{CRx1993}
{Chen} K.,  {Ruderman} M.,  1993, \mn@doi [\apj] {10.1086/172129}, \href
  {https://ui.adsabs.harvard.edu/abs/1993ApJ...402..264C} {402, 264}

\bibitem[\protect\citeauthoryear{{Chen} \& {Wang}}{{Chen} \&
  {Wang}}{2014}]{CWx2014}
{Chen} J.~L.,  {Wang} H.~G.,  2014, \mn@doi [\apjs]
  {10.1088/0067-0049/215/1/11}, \href
  {https://ui.adsabs.harvard.edu/abs/2014ApJS..215...11C} {215, 11}

\bibitem[\protect\citeauthoryear{{Chen}, {Wang}, {Chen}, {Zhang}  \&
  {Liu}}{{Chen} et~al.}{2007}]{CW+2007}
{Chen} J.-L.,  {Wang} H.-G.,  {Chen} W.-H.,  {Zhang} H.,   {Liu} Y.,  2007,
  \mn@doi [\cjaa] {10.1088/1009-9271/7/6/06}, \href
  {https://ui.adsabs.harvard.edu/abs/2007ChJAA...7..789C} {7, 789}

\bibitem[\protect\citeauthoryear{{Cordes}}{{Cordes}}{1978}]{Cxx1978}
{Cordes} J.~M.,  1978, \mn@doi [\apj] {10.1086/156218}, \href
  {https://ui.adsabs.harvard.edu/abs/1978ApJ...222.1006C} {222, 1006}

\bibitem[\protect\citeauthoryear{{Craig}}{{Craig}}{2014}]{Cxx2014}
{Craig} H.~A.,  2014, \mn@doi [\apj] {10.1088/0004-637X/790/2/102}, \href
  {https://ui.adsabs.harvard.edu/abs/2014ApJ...790..102C} {790, 102}

\bibitem[\protect\citeauthoryear{{Craig} \& {Romani}}{{Craig} \&
  {Romani}}{2012}]{CRx2012}
{Craig} H.~A.,  {Romani} R.~W.,  2012, \mn@doi [\apj]
  {10.1088/0004-637X/755/2/137}, \href
  {https://ui.adsabs.harvard.edu/abs/2012ApJ...755..137C} {755, 137}

\bibitem[\protect\citeauthoryear{{Dai} et~al.,}{{Dai} et~al.}{2018}]{DJ+2018}
{Dai} S.,  et~al., 2018, \mn@doi [\mnras] {10.1093/mnras/sty2063}, \href
  {https://ui.adsabs.harvard.edu/abs/2018MNRAS.480.3584D} {480, 3584}

\bibitem[\protect\citeauthoryear{{Dai} et~al.,}{{Dai} et~al.}{2019}]{DL+2019}
{Dai} S.,  et~al., 2019, \mn@doi [\apjl] {10.3847/2041-8213/ab0e7a}, \href
  {https://ui.adsabs.harvard.edu/abs/2019ApJ...874L..14D} {874, L14}

\bibitem[\protect\citeauthoryear{{Desvignes}, {Kramer}, {Cognard}, {Kasian},
  {van Leeuwen}, {Stairs}  \& {Theureau}}{{Desvignes} et~al.}{2013}]{DK+2013}
{Desvignes} G.,  {Kramer} M.,  {Cognard} I.,  {Kasian} L.,  {van Leeuwen} J.,
  {Stairs} I.,   {Theureau} G.,  2013, in {van Leeuwen} J.,  ed.,  IAU
  Symposium Vol. 291, Neutron Stars and Pulsars: Challenges and Opportunities
  after 80 years. pp 199--202 (\mn@eprint {arXiv} {1211.3937}),
  \mn@doi{10.1017/S1743921312023630}

\bibitem[\protect\citeauthoryear{{Dyks} \& {Rudak}}{{Dyks} \&
  {Rudak}}{2012}]{DRx2012}
{Dyks} J.,  {Rudak} B.,  2012, \mn@doi [\mnras]
  {10.1111/j.1365-2966.2011.20265.x}, \href
  {https://ui.adsabs.harvard.edu/abs/2012MNRAS.420.3403D} {420, 3403}

\bibitem[\protect\citeauthoryear{{Dyks} \& {Rudak}}{{Dyks} \&
  {Rudak}}{2013}]{DRx2013}
{Dyks} J.,  {Rudak} B.,  2013, \mn@doi [\mnras] {10.1093/mnras/stt1260}, \href
  {https://ui.adsabs.harvard.edu/abs/2013MNRAS.434.3061D} {434, 3061}

\bibitem[\protect\citeauthoryear{{Dyks} \& {Rudak}}{{Dyks} \&
  {Rudak}}{2015}]{DRx2015}
{Dyks} J.,  {Rudak} B.,  2015, \mn@doi [\mnras] {10.1093/mnras/stu2262}, \href
  {https://ui.adsabs.harvard.edu/abs/2015MNRAS.446.2505D} {446, 2505}

\bibitem[\protect\citeauthoryear{{Dyks}, {Rudak}  \& {Demorest}}{{Dyks}
  et~al.}{2010}]{DRD2010}
{Dyks} J.,  {Rudak} B.,   {Demorest} P.,  2010, \mn@doi [\mnras]
  {10.1111/j.1365-2966.2009.15679.x}, \href
  {https://ui.adsabs.harvard.edu/abs/2010MNRAS.401.1781D} {401, 1781}

\bibitem[\protect\citeauthoryear{{Eatough} et~al.,}{{Eatough}
  et~al.}{2013}]{EF+2013}
{Eatough} R.~P.,  et~al., 2013, \mn@doi [\nat] {10.1038/nature12499}, \href
  {https://ui.adsabs.harvard.edu/abs/2013Natur.501..391E} {501, 391}

\bibitem[\protect\citeauthoryear{{Esposito} et~al.,}{{Esposito}
  et~al.}{2020}]{ER+2020}
{Esposito} P.,  et~al., 2020, \mn@doi [\apjl] {10.3847/2041-8213/ab9742}, \href
  {https://ui.adsabs.harvard.edu/abs/2020ApJ...896L..30E} {896, L30}

\bibitem[\protect\citeauthoryear{{Everett} \& {Weisberg}}{{Everett} \&
  {Weisberg}}{2001}]{EWx2001}
{Everett} J.~E.,  {Weisberg} J.~M.,  2001, \mn@doi [\apj] {10.1086/320652},
  \href {https://ui.adsabs.harvard.edu/abs/2001ApJ...553..341E} {553, 341}

\bibitem[\protect\citeauthoryear{{Gedalin}, {Gruman}  \& {Melrose}}{{Gedalin}
  et~al.}{2002}]{GGM2002}
{Gedalin} M.,  {Gruman} E.,   {Melrose} D.~B.,  2002, \mn@doi [\mnras]
  {10.1046/j.1365-8711.2002.05922.x}, \href
  {https://ui.adsabs.harvard.edu/abs/2002MNRAS.337..422G} {337, 422}

\bibitem[\protect\citeauthoryear{{Geyer} et~al.,}{{Geyer}
  et~al.}{2017}]{GK+2017}
{Geyer} M.,  et~al., 2017, \mn@doi [\mnras] {10.1093/mnras/stx1151}, \href
  {https://ui.adsabs.harvard.edu/abs/2017MNRAS.470.2659G} {470, 2659}

\bibitem[\protect\citeauthoryear{{Gil} \& {Sendyk}}{{Gil} \&
  {Sendyk}}{2000}]{GSx2000}
{Gil} J.~A.,  {Sendyk} M.,  2000, \mn@doi [\apj] {10.1086/309394}, \href
  {https://ui.adsabs.harvard.edu/abs/2000ApJ...541..351G} {541, 351}

\bibitem[\protect\citeauthoryear{{Gil}, {Gronkowski}  \& {Rudnicki}}{{Gil}
  et~al.}{1984}]{GGR1984}
{Gil} J.,  {Gronkowski} P.,   {Rudnicki} W.,  1984, \aap, \href
  {https://ui.adsabs.harvard.edu/abs/1984A&A...132..312G} {132, 312}

\bibitem[\protect\citeauthoryear{{Gil}, {Lyubarsky}  \& {Melikidze}}{{Gil}
  et~al.}{2004}]{GLM2004}
{Gil} J.,  {Lyubarsky} Y.,   {Melikidze} G.~I.,  2004, \mn@doi [\apj]
  {10.1086/379972}, \href
  {https://ui.adsabs.harvard.edu/abs/2004ApJ...600..872G} {600, 872}

\bibitem[\protect\citeauthoryear{{Gotthelf} \& {Halpern}}{{Gotthelf} \&
  {Halpern}}{2007}]{GHx2007}
{Gotthelf} E.~V.,  {Halpern} J.~P.,  2007, \mn@doi [\apss]
  {10.1007/s10509-007-9327-9}, \href
  {https://ui.adsabs.harvard.edu/abs/2007Ap&SS.308...79G} {308, 79}

\bibitem[\protect\citeauthoryear{{Gould} \& {Lyne}}{{Gould} \&
  {Lyne}}{1998}]{GLx1998}
{Gould} D.~M.,  {Lyne} A.~G.,  1998, \mn@doi [\mnras]
  {10.1046/j.1365-8711.1998.02018.x}, \href
  {https://ui.adsabs.harvard.edu/abs/1998MNRAS.301..235G} {301, 235}

\bibitem[\protect\citeauthoryear{{Gupta} \& {Gangadhara}}{{Gupta} \&
  {Gangadhara}}{2003}]{GGx2003}
{Gupta} Y.,  {Gangadhara} R.~T.,  2003, \mn@doi [\apj] {10.1086/345682}, \href
  {https://ui.adsabs.harvard.edu/abs/2003ApJ...584..418G} {584, 418}

\bibitem[\protect\citeauthoryear{{Haberl}}{{Haberl}}{2007}]{Hxx2007}
{Haberl} F.,  2007, \mn@doi [\apss] {10.1007/s10509-007-9342-x}, \href
  {https://ui.adsabs.harvard.edu/abs/2007Ap&SS.308..181H} {308, 181}

\bibitem[\protect\citeauthoryear{{Haslam}, {Salter}, {Stoffel}  \&
  {Wilson}}{{Haslam} et~al.}{1982}]{HS+1982}
{Haslam} C.~G.~T.,  {Salter} C.~J.,  {Stoffel} H.,   {Wilson} W.~E.,  1982,
  \aaps, \href {https://ui.adsabs.harvard.edu/abs/1982A&AS...47....1H} {47, 1}

\bibitem[\protect\citeauthoryear{{Helfand}, {Manchester}  \&
  {Taylor}}{{Helfand} et~al.}{1975}]{HMTx1975}
{Helfand} D.~J.,  {Manchester} R.~N.,   {Taylor} J.~H.,  1975, \mn@doi [\apj]
  {10.1086/153644}, \href
  {https://ui.adsabs.harvard.edu/abs/1975ApJ...198..661H} {198, 661}

\bibitem[\protect\citeauthoryear{{Hibschman} \& {Arons}}{{Hibschman} \&
  {Arons}}{2001}]{HAx2001}
{Hibschman} J.~A.,  {Arons} J.,  2001, \mn@doi [\apj] {10.1086/323069}, \href
  {https://ui.adsabs.harvard.edu/abs/2001ApJ...560..871H} {560, 871}

\bibitem[\protect\citeauthoryear{{Hotan}, {van Straten}  \&
  {Manchester}}{{Hotan} et~al.}{2004}]{psrchive}
{Hotan} A.~W.,  {van Straten} W.,   {Manchester} R.~N.,  2004, \mn@doi [\pasa]
  {10.1071/AS04022}, \href
  {https://ui.adsabs.harvard.edu/abs/2004PASA...21..302H} {21, 302}

\bibitem[\protect\citeauthoryear{{Ilie}, {Johnston}  \& {Weltevrede}}{{Ilie}
  et~al.}{2019}]{IJW2019}
{Ilie} C.~D.,  {Johnston} S.,   {Weltevrede} P.,  2019, \mn@doi [\mnras]
  {10.1093/mnras/sty3315}, \href
  {https://ui.adsabs.harvard.edu/abs/2019MNRAS.483.2778I} {483, 2778}

\bibitem[\protect\citeauthoryear{{Israel} et~al.,}{{Israel}
  et~al.}{2010}]{IE+2010}
{Israel} G.~L.,  et~al., 2010, \mn@doi [\mnras]
  {10.1111/j.1365-2966.2010.17001.x}, \href
  {https://ui.adsabs.harvard.edu/abs/2010MNRAS.408.1387I} {408, 1387}

\bibitem[\protect\citeauthoryear{{Izvekova}, {Kuzmin}, {Malofeev}  \&
  {Shitov}}{{Izvekova} et~al.}{1981}]{IK+1981}
{Izvekova} V.~A.,  {Kuzmin} A.~D.,  {Malofeev} V.~M.,   {Shitov} I.~P.,  1981,
  \mn@doi [\apss] {10.1007/BF00654022}, \href
  {https://ui.adsabs.harvard.edu/abs/1981Ap&SS..78...45I} {78, 45}

\bibitem[\protect\citeauthoryear{{Jankowski}, {van Straten}, {Keane}, {Bailes},
  {Barr}, {Johnston}  \& {Kerr}}{{Jankowski} et~al.}{2018}]{JS+2018}
{Jankowski} F.,  {van Straten} W.,  {Keane} E.~F.,  {Bailes} M.,  {Barr} E.~D.,
   {Johnston} S.,   {Kerr} M.,  2018, \mn@doi [\mnras] {10.1093/mnras/stx2476},
  \href {https://ui.adsabs.harvard.edu/abs/2018MNRAS.473.4436J} {473, 4436}

\bibitem[\protect\citeauthoryear{{Jiang} et~al.,}{{Jiang}
  et~al.}{2020}]{19beam}
{Jiang} P.,  et~al., 2020, \mn@doi [Research in Astronomy and Astrophysics]
  {10.1088/1674-4527/20/5/64}, \href
  {https://ui.adsabs.harvard.edu/abs/2020RAA....20...64J} {20, 064}

\bibitem[\protect\citeauthoryear{{Johnston} \& {Karastergiou}}{{Johnston} \&
  {Karastergiou}}{2019}]{JKx2019}
{Johnston} S.,  {Karastergiou} A.,  2019, \mn@doi [\mnras]
  {10.1093/mnras/stz400}, \href
  {https://ui.adsabs.harvard.edu/abs/2019MNRAS.485..640J} {485, 640}

\bibitem[\protect\citeauthoryear{{Johnston}, {Smith}, {Karastergiou}  \&
  {Kramer}}{{Johnston} et~al.}{2020}]{JSK2020}
{Johnston} S.,  {Smith} D.~A.,  {Karastergiou} A.,   {Kramer} M.,  2020,
  \mn@doi [\mnras] {10.1093/mnras/staa2110}, \href
  {https://ui.adsabs.harvard.edu/abs/2020MNRAS.497.1957J} {497, 1957}

\bibitem[\protect\citeauthoryear{{Karastergiou} \& {Johnston}}{{Karastergiou}
  \& {Johnston}}{2007a}]{KJx2007}
{Karastergiou} A.,  {Johnston} S.,  2007a, \mn@doi [\mnras]
  {10.1111/j.1365-2966.2007.12237.x}, \href
  {https://ui.adsabs.harvard.edu/abs/2007MNRAS.380.1678K} {380, 1678}

\bibitem[\protect\citeauthoryear{{Karastergiou} \& {Johnston}}{{Karastergiou}
  \& {Johnston}}{2007b}]{KSx2007}
{Karastergiou} A.,  {Johnston} S.,  2007b, \mn@doi [\mnras]
  {10.1111/j.1365-2966.2007.12237.x}, \href
  {https://ui.adsabs.harvard.edu/abs/2007MNRAS.380.1678K} {380, 1678}

\bibitem[\protect\citeauthoryear{{Keith}, {Johnston}, {Weltevrede}  \&
  {Kramer}}{{Keith} et~al.}{2010}]{KJ+2010}
{Keith} M.~J.,  {Johnston} S.,  {Weltevrede} P.,   {Kramer} M.,  2010, \mn@doi
  [\mnras] {10.1111/j.1365-2966.2009.15926.x}, \href
  {https://ui.adsabs.harvard.edu/abs/2010MNRAS.402..745K} {402, 745}

\bibitem[\protect\citeauthoryear{{Keith}, {Johnston}, {Levin}  \&
  {Bailes}}{{Keith} et~al.}{2011}]{KJ+2011}
{Keith} M.~J.,  {Johnston} S.,  {Levin} L.,   {Bailes} M.,  2011, \mn@doi
  [\mnras] {10.1111/j.1365-2966.2011.19041.x}, \href
  {https://ui.adsabs.harvard.edu/abs/2011MNRAS.416..346K} {416, 346}

\bibitem[\protect\citeauthoryear{{Kijak} \& {Gil}}{{Kijak} \&
  {Gil}}{2003}]{KGx2003}
{Kijak} J.,  {Gil} J.,  2003, \mn@doi [A\&A] {10.1051/0004-6361:20021583},
  \href {https://ui.adsabs.harvard.edu/abs/2003A&A...397..969K} {397, 969}

\bibitem[\protect\citeauthoryear{{Komesaroff}}{{Komesaroff}}{1970}]{Kxx1970}
{Komesaroff} M.~M.,  1970, \mn@doi [\nat] {10.1038/225612a0}, \href
  {https://ui.adsabs.harvard.edu/abs/1970Natur.225..612K} {225, 612}

\bibitem[\protect\citeauthoryear{{Kramer}, {Stappers}, {Jessner}, {Lyne}  \&
  {Jordan}}{{Kramer} et~al.}{2007}]{KS+2007}
{Kramer} M.,  {Stappers} B.~W.,  {Jessner} A.,  {Lyne} A.~G.,   {Jordan} C.~A.,
   2007, \mn@doi [\mnras] {10.1111/j.1365-2966.2007.11622.x}, \href
  {https://ui.adsabs.harvard.edu/abs/2007MNRAS.377..107K} {377, 107}

\bibitem[\protect\citeauthoryear{{Lawson}, {Mayer}, {Osborne}  \&
  {Parkinson}}{{Lawson} et~al.}{1987}]{LM+1987}
{Lawson} K.~D.,  {Mayer} C.~J.,  {Osborne} J.~L.,   {Parkinson} M.~L.,  1987,
  \mn@doi [\mnras] {10.1093/mnras/225.2.307}, \href
  {https://ui.adsabs.harvard.edu/abs/1987MNRAS.225..307L} {225, 307}

\bibitem[\protect\citeauthoryear{{Lazarus} et~al.,}{{Lazarus}
  et~al.}{2015}]{LB+2015}
{Lazarus} P.,  et~al., 2015, \mn@doi [\apj] {10.1088/0004-637X/812/1/81}, \href
  {https://ui.adsabs.harvard.edu/abs/2015ApJ...812...81L} {812, 81}

\bibitem[\protect\citeauthoryear{{Lazarus}, {Karuppusamy}, {Graikou},
  {Caballero}, {Champion}, {Lee}, {Verbiest}  \& {Kramer}}{{Lazarus}
  et~al.}{2016}]{coastguard}
{Lazarus} P.,  {Karuppusamy} R.,  {Graikou} E.,  {Caballero} R.~N.,  {Champion}
  D.~J.,  {Lee} K.~J.,  {Verbiest} J.~P.~W.,   {Kramer} M.,  2016, \mn@doi
  [\mnras] {10.1093/mnras/stw189}, \href
  {https://ui.adsabs.harvard.edu/abs/2016MNRAS.458..868L} {458, 868}

\bibitem[\protect\citeauthoryear{{Levin} et~al.,}{{Levin}
  et~al.}{2010}]{LB+2010}
{Levin} L.,  et~al., 2010, \mn@doi [\apjl] {10.1088/2041-8205/721/1/L33}, \href
  {https://ui.adsabs.harvard.edu/abs/2010ApJ...721L..33L} {721, L33}

\bibitem[\protect\citeauthoryear{{Levin} et~al.,}{{Levin}
  et~al.}{2012}]{LB+2012}
{Levin} L.,  et~al., 2012, \mn@doi [\mnras] {10.1111/j.1365-2966.2012.20807.x},
  \href {https://ui.adsabs.harvard.edu/abs/2012MNRAS.422.2489L} {422, 2489}

\bibitem[\protect\citeauthoryear{{Levin} et~al.,}{{Levin}
  et~al.}{2019}]{LL+2019}
{Levin} L.,  et~al., 2019, \mn@doi [\mnras] {10.1093/mnras/stz2074}, \href
  {https://ui.adsabs.harvard.edu/abs/2019MNRAS.488.5251L} {488, 5251}

\bibitem[\protect\citeauthoryear{{Li} et~al.,}{{Li} et~al.}{2018}]{LW+2018}
{Li} D.,  et~al., 2018, \mn@doi [IEEE Microwave Magazine]
  {10.1109/MMM.2018.2802178}, \href
  {https://ui.adsabs.harvard.edu/abs/2018IMMag..19..112L} {19, 112}

\bibitem[\protect\citeauthoryear{{Liu}, {Keane}, {Lee}, {Kramer}, {Cordes}  \&
  {Purver}}{{Liu} et~al.}{2012}]{LKL+2012}
{Liu} K.,  {Keane} E.~F.,  {Lee} K.~J.,  {Kramer} M.,  {Cordes} J.~M.,
  {Purver} M.~B.,  2012, \mn@doi [\mnras] {10.1111/j.1365-2966.2011.20041.x},
  \href {https://ui.adsabs.harvard.edu/abs/2012MNRAS.420..361L} {420, 361}

\bibitem[\protect\citeauthoryear{{Lorimer} \& {Kramer}}{{Lorimer} \&
  {Kramer}}{2005}]{handbook}
{Lorimer} D.~R.,  {Kramer} M.,  2005, Handbook of Pulsar Astronomy.
Cambridge University Press

\bibitem[\protect\citeauthoryear{{Lower}, {Shannon}, {Johnston}  \&
  {Bailes}}{{Lower} et~al.}{2020}]{LS+2020}
{Lower} M.~E.,  {Shannon} R.~M.,  {Johnston} S.,   {Bailes} M.,  2020, \mn@doi
  [\apjl] {10.3847/2041-8213/ab9898}, \href
  {https://ui.adsabs.harvard.edu/abs/2020ApJ...896L..37L} {896, L37}

\bibitem[\protect\citeauthoryear{{Lower}, {Johnston}, {Shannon}, {Bailes}  \&
  {Camilo}}{{Lower} et~al.}{2021}]{LJS+2021}
{Lower} M.~E.,  {Johnston} S.,  {Shannon} R.~M.,  {Bailes} M.,   {Camilo} F.,
  2021, \mn@doi [\mnras] {10.1093/mnras/staa3789}, \href
  {https://ui.adsabs.harvard.edu/abs/2021MNRAS.502..127L} {502, 127}

\bibitem[\protect\citeauthoryear{{Lyne} \& {Graham-Smith}}{{Lyne} \&
  {Graham-Smith}}{2012}]{PulsarAstronomy}
{Lyne} A.,  {Graham-Smith} F.,  2012, Pulsar Astronomy, 4 edn.
Cambridge University Press

\bibitem[\protect\citeauthoryear{{Lyne} \& {Manchester}}{{Lyne} \&
  {Manchester}}{1988}]{LMx1988}
{Lyne} A.~G.,  {Manchester} R.~N.,  1988, \mn@doi [\mnras]
  {10.1093/mnras/234.3.477}, \href
  {https://ui.adsabs.harvard.edu/abs/1988MNRAS.234..477L} {234, 477}

\bibitem[\protect\citeauthoryear{{Malofeev} \& {Malov}}{{Malofeev} \&
  {Malov}}{1980}]{MMx1980}
{Malofeev} V.~M.,  {Malov} I.~F.,  1980, \sovast, \href
  {https://ui.adsabs.harvard.edu/abs/1980SvA....24...54M} {24, 54}

\bibitem[\protect\citeauthoryear{{Manchester} et~al.,}{{Manchester}
  et~al.}{2010}]{MK+2010}
{Manchester} R.~N.,  et~al., 2010, \mn@doi [\apj]
  {10.1088/0004-637X/710/2/1694}, \href
  {https://ui.adsabs.harvard.edu/abs/2010ApJ...710.1694M} {710, 1694}

\bibitem[\protect\citeauthoryear{{McKinnon} \& {Stinebring}}{{McKinnon} \&
  {Stinebring}}{2000}]{MSx2000}
{McKinnon} M.~M.,  {Stinebring} D.~R.,  2000, \mn@doi [\apj] {10.1086/308264},
  \href {https://ui.adsabs.harvard.edu/abs/2000ApJ...529..435M} {529, 435}

\bibitem[\protect\citeauthoryear{{Michel}}{{Michel}}{1987}]{Mxx1987}
{Michel} F.~C.,  1987, \mn@doi [\apj] {10.1086/165775}, \href
  {https://ui.adsabs.harvard.edu/abs/1987ApJ...322..822M} {322, 822}

\bibitem[\protect\citeauthoryear{{Mitra} \& {Rankin}}{{Mitra} \&
  {Rankin}}{2002}]{MRx2002}
{Mitra} D.,  {Rankin} J.~M.,  2002, \mn@doi [\apj] {10.1086/342136}, \href
  {https://ui.adsabs.harvard.edu/abs/2002ApJ...577..322M} {577, 322}

\bibitem[\protect\citeauthoryear{{Mitra}, {Basu}, {Melikidze}  \&
  {Arjunwadkar}}{{Mitra} et~al.}{2020}]{MB+2020}
{Mitra} D.,  {Basu} R.,  {Melikidze} G.~I.,   {Arjunwadkar} M.,  2020, \mn@doi
  [\mnras] {10.1093/mnras/stz3620}, \href
  {https://ui.adsabs.harvard.edu/abs/2020MNRAS.492.2468M} {492, 2468}

\bibitem[\protect\citeauthoryear{{Morello} et~al.,}{{Morello}
  et~al.}{2020}]{MK+2020}
{Morello} V.,  et~al., 2020, \mn@doi [\mnras] {10.1093/mnras/staa321}, \href
  {https://ui.adsabs.harvard.edu/abs/2020MNRAS.493.1165M} {493, 1165}

\bibitem[\protect\citeauthoryear{{Nan} et~al.,}{{Nan} et~al.}{2011}]{FASTpaper}
{Nan} R.,  et~al., 2011, \mn@doi [International Journal of Modern Physics D]
  {10.1142/S0218271811019335}, \href
  {https://ui.adsabs.harvard.edu/abs/2011IJMPD..20..989N} {20, 989}

\bibitem[\protect\citeauthoryear{{Perna} \& {Gotthelf}}{{Perna} \&
  {Gotthelf}}{2008}]{PGx2008}
{Perna} R.,  {Gotthelf} E.~V.,  2008, \mn@doi [\apj] {10.1086/588211}, \href
  {https://ui.adsabs.harvard.edu/abs/2008ApJ...681..522P} {681, 522}

\bibitem[\protect\citeauthoryear{{Pilia} et~al.,}{{Pilia}
  et~al.}{2016}]{PH+2016}
{Pilia} M.,  et~al., 2016, \mn@doi [\aap] {10.1051/0004-6361/201425196}, \href
  {https://ui.adsabs.harvard.edu/abs/2016A&A...586A..92P} {586, A92}

\bibitem[\protect\citeauthoryear{{Radhakrishnan} \& {Cooke}}{{Radhakrishnan} \&
  {Cooke}}{1969}]{RCx1969}
{Radhakrishnan} V.,  {Cooke} D.~J.,  1969, \aplett, \href
  {https://ui.adsabs.harvard.edu/abs/1969ApL.....3..225R} {3, 225}

\bibitem[\protect\citeauthoryear{{Radhakrishnan} \& {Rankin}}{{Radhakrishnan}
  \& {Rankin}}{1990}]{RRx1990}
{Radhakrishnan} V.,  {Rankin} J.~M.,  1990, \mn@doi [\apj] {10.1086/168531},
  \href {https://ui.adsabs.harvard.edu/abs/1990ApJ...352..258R} {352, 258}

\bibitem[\protect\citeauthoryear{{Rankin}}{{Rankin}}{1983a}]{Rxx1983a}
{Rankin} J.~M.,  1983a, \mn@doi [\apj] {10.1086/161450}, \href
  {https://ui.adsabs.harvard.edu/abs/1983ApJ...274..333R} {274, 333}

\bibitem[\protect\citeauthoryear{{Rankin}}{{Rankin}}{1983b}]{Rxx1983b}
{Rankin} J.~M.,  1983b, \mn@doi [\apj] {10.1086/161451}, \href
  {https://ui.adsabs.harvard.edu/abs/1983ApJ...274..359R} {274, 359}

\bibitem[\protect\citeauthoryear{{Rankin}}{{Rankin}}{1986}]{Rxx1986}
{Rankin} J.~M.,  1986, \mn@doi [\apj] {10.1086/163955}, \href
  {https://ui.adsabs.harvard.edu/abs/1986ApJ...301..901R} {301, 901}

\bibitem[\protect\citeauthoryear{{Rankin}}{{Rankin}}{1990}]{Rxx1990}
{Rankin} J.~M.,  1990, \mn@doi [\apj] {10.1086/168530}, \href
  {https://ui.adsabs.harvard.edu/abs/1990ApJ...352..247R} {352, 247}

\bibitem[\protect\citeauthoryear{{Rankin}}{{Rankin}}{1993}]{Rxx1993}
{Rankin} J.~M.,  1993, \mn@doi [\apj] {10.1086/172361}, \href
  {https://ui.adsabs.harvard.edu/abs/1993ApJ...405..285R} {405, 285}

\bibitem[\protect\citeauthoryear{{Rankin}, {Stinebring}  \&
  {Weisberg}}{{Rankin} et~al.}{1989}]{RSW1989}
{Rankin} J.~M.,  {Stinebring} D.~R.,   {Weisberg} J.~M.,  1989, \mn@doi [\apj]
  {10.1086/168068}, \href
  {https://ui.adsabs.harvard.edu/abs/1989ApJ...346..869R} {346, 869}

\bibitem[\protect\citeauthoryear{{Rankin}, {Olszanski}  \& {Wright}}{{Rankin}
  et~al.}{2020}]{ROWx2020}
{Rankin} J.~M.,  {Olszanski} T. E.~E.,   {Wright} G. A.~E.,  2020, \mn@doi
  [\apj] {10.3847/1538-4357/ab67cb}, \href
  {https://ui.adsabs.harvard.edu/abs/2020ApJ...890..151R} {890, 151}

\bibitem[\protect\citeauthoryear{{Rathnasree} \& {Rankin}}{{Rathnasree} \&
  {Rankin}}{1995}]{RRxx1995}
{Rathnasree} N.,  {Rankin} J.~M.,  1995, \mn@doi [\apj] {10.1086/176349}, \href
  {https://ui.adsabs.harvard.edu/abs/1995ApJ...452..814R} {452, 814}

\bibitem[\protect\citeauthoryear{{Reich} \& {Reich}}{{Reich} \&
  {Reich}}{1988}]{RRx1988}
{Reich} P.,  {Reich} W.,  1988, \aap, \href
  {https://ui.adsabs.harvard.edu/abs/1988A&A...196..211R} {196, 211}

\bibitem[\protect\citeauthoryear{{Rookyard}, {Weltevrede}  \&
  {Johnston}}{{Rookyard} et~al.}{2015a}]{RWJ2015b}
{Rookyard} S.~C.,  {Weltevrede} P.,   {Johnston} S.,  2015a, \mn@doi [\mnras]
  {10.1093/mnras/stu2083}, \href
  {https://ui.adsabs.harvard.edu/abs/2015MNRAS.446.3356R} {446, 3356}

\bibitem[\protect\citeauthoryear{{Rookyard}, {Weltevrede}  \&
  {Johnston}}{{Rookyard} et~al.}{2015b}]{RWJ2015a}
{Rookyard} S.~C.,  {Weltevrede} P.,   {Johnston} S.,  2015b, \mn@doi [\mnras]
  {10.1093/mnras/stu2236}, \href
  {https://ui.adsabs.harvard.edu/abs/2015MNRAS.446.3367R} {446, 3367}

\bibitem[\protect\citeauthoryear{{Ruderman} \& {Sutherland}}{{Ruderman} \&
  {Sutherland}}{1975}]{RSx1975}
{Ruderman} M.~A.,  {Sutherland} P.~G.,  1975, \mn@doi [\apj] {10.1086/153393},
  \href {https://ui.adsabs.harvard.edu/abs/1975ApJ...196...51R} {196, 51}

\bibitem[\protect\citeauthoryear{{Sanidas} et~al.,}{{Sanidas}
  et~al.}{2019}]{SC+2019}
{Sanidas} S.,  et~al., 2019, \mn@doi [\aap] {10.1051/0004-6361/201935609},
  \href {https://ui.adsabs.harvard.edu/abs/2019A&A...626A.104S} {626, A104}

\bibitem[\protect\citeauthoryear{{Serylak} et~al.,}{{Serylak}
  et~al.}{2009}]{SS+2009}
{Serylak} M.,  et~al., 2009, \mn@doi [\mnras]
  {10.1111/j.1365-2966.2008.14260.x}, \href
  {https://ui.adsabs.harvard.edu/abs/2009MNRAS.394..295S} {394, 295}

\bibitem[\protect\citeauthoryear{{Shimwell} et~al.,}{{Shimwell}
  et~al.}{2017}]{SR+2017}
{Shimwell} T.~W.,  et~al., 2017, \mn@doi [\aap] {10.1051/0004-6361/201629313},
  \href {https://ui.adsabs.harvard.edu/abs/2017A&A...598A.104S} {598, A104}

\bibitem[\protect\citeauthoryear{{Sieber}}{{Sieber}}{1973}]{Sxx1973}
{Sieber} W.,  1973, \aap, \href
  {https://ui.adsabs.harvard.edu/abs/1973A&A....28..237S} {28, 237}

\bibitem[\protect\citeauthoryear{{Slee}, {Dulk}  \& {Otrupcek}}{{Slee}
  et~al.}{1980}]{SDO1980}
{Slee} O.~B.,  {Dulk} G.~A.,   {Otrupcek} R.~E.,  1980, Proceedings of the
  Astronomical Society of Australia, \href
  {https://ui.adsabs.harvard.edu/abs/1980PASAu...4..100S} {4, 100}

\bibitem[\protect\citeauthoryear{{Sobey} et~al.,}{{Sobey}
  et~al.}{2019}]{SB+2019}
{Sobey} C.,  et~al., 2019, \mn@doi [\mnras] {10.1093/mnras/stz214}, \href
  {https://ui.adsabs.harvard.edu/abs/2019MNRAS.484.3646S} {484, 3646}

\bibitem[\protect\citeauthoryear{{Spitkovsky}}{{Spitkovsky}}{2006}]{Sxx2006}
{Spitkovsky} A.,  2006, \mn@doi [\apjl] {10.1086/507518}, \href
  {https://ui.adsabs.harvard.edu/abs/2006ApJ...648L..51S} {648, L51}

\bibitem[\protect\citeauthoryear{{Stappers} et~al.,}{{Stappers}
  et~al.}{2011}]{SH+2011}
{Stappers} B.~W.,  et~al., 2011, \mn@doi [\aap] {10.1051/0004-6361/201116681},
  \href {https://ui.adsabs.harvard.edu/abs/2011A&A...530A..80S} {530, A80}

\bibitem[\protect\citeauthoryear{{Sturrock}}{{Sturrock}}{1971}]{Sxx1971}
{Sturrock} P.~A.,  1971, \mn@doi [\apj] {10.1086/150865}, \href
  {https://ui.adsabs.harvard.edu/abs/1971ApJ...164..529S} {164, 529}

\bibitem[\protect\citeauthoryear{{Szary}}{{Szary}}{2013}]{Sxxx2013}
{Szary} A.,  2013, PhD thesis, University of Zielona G\'ora, \url
  {https://arxiv.org/pdf/1304.4203.pdf}

\bibitem[\protect\citeauthoryear{{Tan} et~al.,}{{Tan} et~al.}{2018}]{TB+2018}
{Tan} C.~M.,  et~al., 2018, \mn@doi [\apj] {10.3847/1538-4357/aade88}, \href
  {https://ui.adsabs.harvard.edu/abs/2018ApJ...866...54T} {866, 54}

\bibitem[\protect\citeauthoryear{{Thorsett}}{{Thorsett}}{1991}]{Txx1991}
{Thorsett} S.~E.,  1991, \mn@doi [\apj] {10.1086/170355}, \href
  {https://ui.adsabs.harvard.edu/abs/1991ApJ...377..263T} {377, 263}

\bibitem[\protect\citeauthoryear{{Timokhin}}{{Timokhin}}{2010}]{Txx2010}
{Timokhin} A.~N.,  2010, \mn@doi [\mnras] {10.1111/j.1365-2966.2010.17286.x},
  \href {https://ui.adsabs.harvard.edu/abs/2010MNRAS.408.2092T} {408, 2092}

\bibitem[\protect\citeauthoryear{{Timokhin} \& {Harding}}{{Timokhin} \&
  {Harding}}{2015}]{THxx2015}
{Timokhin} A.~N.,  {Harding} A.~K.,  2015, \mn@doi [\apj]
  {10.1088/0004-637X/810/2/144}, \href
  {https://ui.adsabs.harvard.edu/abs/2015ApJ...810..144T} {810, 144}

\bibitem[\protect\citeauthoryear{{V{\'e}ron-Cetty} \&
  {V{\'e}ron}}{{V{\'e}ron-Cetty} \& {V{\'e}ron}}{2006}]{VVx2006}
{V{\'e}ron-Cetty} M.~P.,  {V{\'e}ron} P.,  2006, \mn@doi [\aap]
  {10.1051/0004-6361:20065177}, \href
  {https://ui.adsabs.harvard.edu/abs/2006A&A...455..773V} {455, 773}

\bibitem[\protect\citeauthoryear{{Vigan{\`o}}, {Rea}, {Pons}, {Perna},
  {Aguilera}  \& {Miralles}}{{Vigan{\`o}} et~al.}{2013}]{VR+2013}
{Vigan{\`o}} D.,  {Rea} N.,  {Pons} J.~A.,  {Perna} R.,  {Aguilera} D.~N.,
  {Miralles} J.~A.,  2013, \mn@doi [\mnras] {10.1093/mnras/stt1008}, \href
  {https://ui.adsabs.harvard.edu/abs/2013MNRAS.434..123V} {434, 123}

\bibitem[\protect\citeauthoryear{{Wang} et~al.,}{{Wang} et~al.}{2014}]{WP+2014}
{Wang} H.~G.,  et~al., 2014, \mn@doi [\apj] {10.1088/0004-637X/789/1/73}, \href
  {https://ui.adsabs.harvard.edu/abs/2014ApJ...789...73W} {789, 73}

\bibitem[\protect\citeauthoryear{{Wardle} \& {Kronberg}}{{Wardle} \&
  {Kronberg}}{1974}]{WKx1974}
{Wardle} J.~F.~C.,  {Kronberg} P.~P.,  1974, \mn@doi [\apj] {10.1086/153240},
  \href {https://ui.adsabs.harvard.edu/abs/1974ApJ...194..249W} {194, 249}

\bibitem[\protect\citeauthoryear{{Weltevrede}}{{Weltevrede}}{2016}]{psrsalsa}
{Weltevrede} P.,  2016, \mn@doi [\aap] {10.1051/0004-6361/201527950}, \href
  {https://ui.adsabs.harvard.edu/abs/2016A&A...590A.109W} {590, A109}

\bibitem[\protect\citeauthoryear{{Weltevrede} \& {Johnston}}{{Weltevrede} \&
  {Johnston}}{2008}]{WJx2008b}
{Weltevrede} P.,  {Johnston} S.,  2008, \mn@doi [\mnras]
  {10.1111/j.1365-2966.2008.13950.x}, \href
  {https://ui.adsabs.harvard.edu/abs/2008MNRAS.391.1210W} {391, 1210}

\bibitem[\protect\citeauthoryear{{Young}, {Manchester}  \& {Johnston}}{{Young}
  et~al.}{1999}]{YMJ1999}
{Young} M.~D.,  {Manchester} R.~N.,   {Johnston} S.,  1999, \mn@doi [\nat]
  {10.1038/23650}, \href
  {https://ui.adsabs.harvard.edu/abs/1999Natur.400..848Y} {400, 848}

\bibitem[\protect\citeauthoryear{{Zarka}, {Denis}, {Tagger}  et~al.}{{Zarka}
  et~al.}{2020}]{ZD+2020}
{Zarka} P.,  {Denis} L.,  {Tagger} M.,   et~al., 2020, URSI GASS 2020, Session
  J01 New Telescopes on the Frontier

\bibitem[\protect\citeauthoryear{{Zhang}, {Harding}  \& {Muslimov}}{{Zhang}
  et~al.}{2000}]{ZHM2000}
{Zhang} B.,  {Harding} A.~K.,   {Muslimov} A.~G.,  2000, \mn@doi [\apjl]
  {10.1086/312542}, \href
  {https://ui.adsabs.harvard.edu/abs/2000ApJ...531L.135Z} {531, L135}

\bibitem[\protect\citeauthoryear{{van Haarlem} et~al.,}{{van Haarlem}
  et~al.}{2013}]{lofarPaper}
{van Haarlem} M.~P.,  et~al., 2013, \mn@doi [\aap]
  {10.1051/0004-6361/201220873}, \href
  {https://ui.adsabs.harvard.edu/abs/2013A&A...556A...2V} {556, A2}

\bibitem[\protect\citeauthoryear{{van Heerden}, {Karastergiou}  \&
  {Roberts}}{{van Heerden} et~al.}{2017}]{HKR2017}
{van Heerden} E.,  {Karastergiou} A.,   {Roberts} S.~J.,  2017, \mn@doi
  [\mnras] {10.1093/mnras/stw3068}, \href
  {https://ui.adsabs.harvard.edu/abs/2017MNRAS.467.1661V} {467, 1661}

\bibitem[\protect\citeauthoryear{{van Kerkwijk} \& {Kaplan}}{{van Kerkwijk} \&
  {Kaplan}}{2007}]{KKx2007}
{van Kerkwijk} M.~H.,  {Kaplan} D.~L.,  2007, \mn@doi [\apss]
  {10.1007/s10509-007-9343-9}, \href
  {https://ui.adsabs.harvard.edu/abs/2007Ap&SS.308..191V} {308, 191}

\bibitem[\protect\citeauthoryear{{van Straten} \& {Bailes}}{{van Straten} \&
  {Bailes}}{2011}]{SBx2011}
{van Straten} W.,  {Bailes} M.,  2011, \mn@doi [\pasa] {10.1071/AS10021}, \href
  {https://ui.adsabs.harvard.edu/abs/2011PASA...28....1V} {28, 1}

\makeatother
\end{thebibliography}


\appendix

\section{Constraints on the viewing geometry}
\label{app: expanded geometry derivation}

This appendix gives a more detailed derivation of the viewing geometry based on fitting the RVM and considering the observed pulse width, which was summarised in Sec.~\ref{sec: polarisation and geometry}. The RVM describes the shape of the position angle (PA; $\psi$) curve as a function of pulse longitude, $\phi$. It depends on the magnetic inclination angle, $\alpha$ and the impact parameter of the observer's line of sight, $\beta$, and can be expressed as 
\begin{equation}
	\label{eq: RVM}
		\Delta\psi = \arctan\bigg(  \frac{\sin(\Delta\phi) \sin\alpha }{\sin\zeta\cos\alpha - \cos\zeta\sin\alpha\cos(\Delta\phi) }  \bigg),
\end{equation}
where $\Delta\psi = \psi - \psi_0$, $\Delta\phi = \phi - \phi_0$, and $\zeta = \alpha + \beta$. It describes a monotonic S-shaped curve which has an inflection point at $(\phi_0,\ \psi_0)$.

\begin{figure}
 \includegraphics[width=\columnwidth]{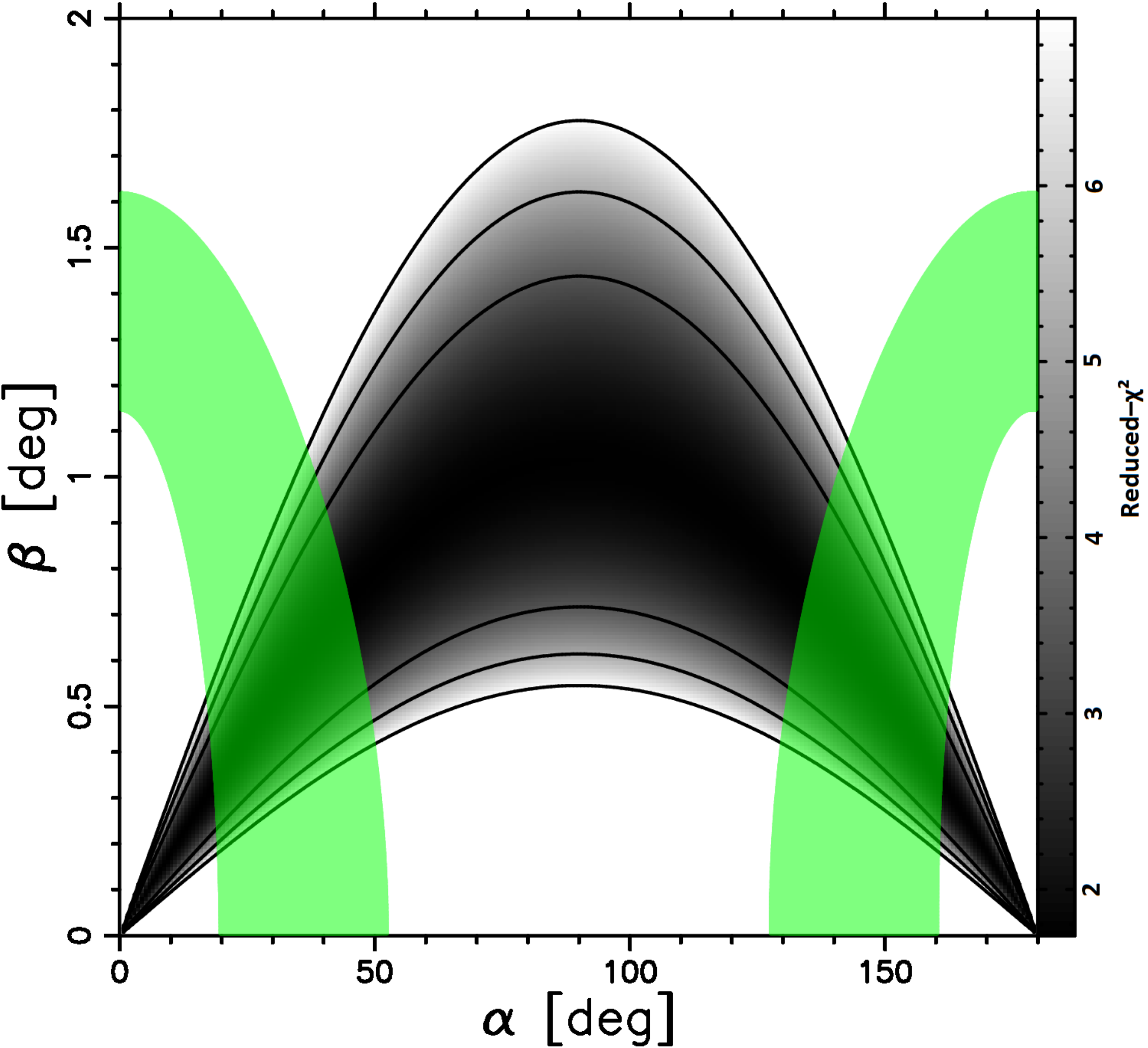}
 \caption{The goodness-of-fit (reduced-$\chi^2$) of the RVM to the PA curve as a function of ($\alpha$, $\beta$) space obtained for the LOFAR Core data is shown in grey-scale. The black contours correspond to a reduced-$\chi^2$ of two-, three-, and four-times the minimum value. The green shaded regions are the ``allowed'' viewing geometries, which are constraints arising from the estimated emission height and observed profile width (see the text in Appendix~\ref{app: expanded geometry derivation} for details including the assumptions made).}
 \label{fig: banana}
\end{figure}

The goodness-of-fit of Eq.~\eqref{eq: RVM} to the observed $\psi(\phi)$ is parametrised by the reduced-$\chi^2$ and its variation is shown in Fig.~\ref{fig: banana} for the LOFAR Core data (see \citealt{RWJ2015a} for details of the methodology used). The darker shading corresponds to lower reduced-$\chi^2$ values and so a better fit. The black contours indicate $1\sigma$, $2\sigma$ and $3\sigma$ confidence intervals. As can be expected for a pulsar with a very small duty-cycle, $\alpha$ and $\beta$ are highly correlated. The fit confirms that $\beta$ must be small ($<1.8\degr$), however the magnetic inclination $\alpha$ is unconstrained from RVM fitting alone.

The measured profile widths provide additional information about the opening angle of the radio beam, how the line of sight cuts it, and the emission height. We assume that all radiation of a given frequency is produced at some height $h_\mathrm{em}$ in the magnetosphere in a circular region surrounding the magnetic axis. The emission beam is delimited by tangents to the last open field lines, forming a conal beam. In the small angle limit \citep[$h_\mathrm{em} \ll R_\mathrm{LC}$, see for example][]{Rxx1990} the half opening angle of the emission cone is
\begin{equation}
\label{eq: cone angle}
    \rho = \sqrt{\frac{9\pi h_\mathrm{em}}{2cP}}.
\end{equation}
This implies that the radio beam should widen with increasing emission height, and longer period pulsars can be expected to have narrower beams.

The width of the pulse profile depends on how the line of sight cuts through the emission beam. \citet{GGR1984} showed that the rotational phase range for which the line of sight samples the open-field-line-region, $W$, can be expressed as
\begin{equation}
\label{eq: allowed geometry}
    \cos\rho = \cos\alpha\cos(\alpha+\beta)+\sin\alpha\sin(\alpha+\beta)\cos\bigg(\frac{W}{2}\bigg).
\end{equation}
This means that a measurement of $W$ can help to constrain the parameters $\alpha$ and $\beta$, as well as $h_\mathrm{em}$ via $\rho$ \citep[see for example][]{RWJ2015a}. Here it is important to note that the open-field-line region does not necessarily emit over its full extent, hence the measured profile width does not necessarily correspond to $W$ as defined in Eq.~\eqref{eq: allowed geometry}.

The FAST profile is likely to correspond to a more fully illuminated beam (see Sec.~\ref{sec: discuss width}). Therefore, Fig.~\ref{fig: banana} highlights the geometries which are compatible with the observed pulse width for that observation shown as the green shaded region. Here $W_{10}=4.3\pm0.2\degr$ is used, the width of the profile as defined at 10~per~cent of the peak flux density, to ensure that most emission is from the open field line region. Here it is assumed that the emission height lies within the range of 200 to 400~km \citep[e.g.][]{MRx2002, JKx2019}. Moreover, $W$ as defined in Eq.~\eqref{eq: allowed geometry} is assumed to be between the measured $W_{10}$ and twice the distance between the PA curve inflection point and the furthest edge of the FAST pulse profile, in order to account for potential underfilling of the radio beam (see Sec.~\ref{sec: discuss geometry} for the motivation). Here we take the PA inflection point to coincide with the position of the fiducial plane (the plane containing the magnetic and rotation axes), because the emission height at FAST frequencies is argued to be low enough to make any A/R effects small (see Sec.~\ref{sec: discuss geometry}). Taking into account the uncertainties on $W_{10}$ and the inflection point, this results in $4.1\degr \leq W \leq 6.9\degr$ assuming that at least one edge of the profile corresponds to the boundary with the last open field line region. This allowed range of $W$ and emission height results in a collection of contours in ($\alpha$, $\beta$) space defined by Eqs.~\eqref{eq: cone angle}~and~\eqref{eq: allowed geometry}. These contours (green shaded regions Fig.~\ref{fig: banana}) show that $\beta$ is likely $\leq1.1\degr$, and suggests that the pulsar is relatively aligned (a small $\alpha$). Further to these considerations, it cannot be ruled out that neither edge of the profile reaches the edge of the open field line region. This would correspond to a lower filling fraction and modestly more aligned $\alpha$.

\section{The core-cone model}
\label{app: corecone model consistency}

The profile width evolution of PSR~J0250+5854 can be described by the core-cone model, and the geometry derived in this way is consistent with our results in Sec.~\ref{sec: polarisation and geometry} and Appendix~\ref{app: expanded geometry derivation}. In this scenario, the LOFAR Core profile is associated with emission from the core beam, while the FAST profile is broader due to presence of emission from the inner cone, in the form of ``conal outriders''. The width of the core beam is believed to reflect the size of the polar cap at the surface of the pulsar, and is largely independent of observing frequency \citep{Rxx1983a} and follows the empirical relation $W_{50} = 2.45\degr P^{-0.5} / \sin\alpha$ \citep{Rxx1990}.
For PSR~J0250+5854, $P=23.5$~s and the measured width of the LOFAR Core profile $W_{50} \approx 1\degr$ gives $\alpha \approx 30\degr$. The gradient of the PA curve at the inflection point $\sin\alpha/\sin\beta \approx 55$~deg~deg$^{-1}$ which then implies $\beta \approx 0.5\degr$. These values of $\alpha$ and $\beta$ lie within the ``allowed geometries'' in Fig.~\ref{fig: banana}, which were derived without a specific model in mind. 

Extending this further, \citet{Rxx1993} argued that the half opening angle of the inner cone follows the relation $\rho_\mathrm{inner} = 4.3\degr P^{-0.52}$, which for PSR~J0250+5854 gives $\rho_\mathrm{inner} = 0.83\degr$. Substituting this and the derived $\alpha$ and $\beta$ values into Eq.~\eqref{eq: allowed geometry} gives the expected width of the profile due to the inner cone, $W \approx 2.5\degr$. This is consistent with $W_{50} = 2.4\pm 0.1\degr$ measured for the FAST profile.


\bsp	
\label{lastpage}
\end{document}